\documentclass[twocolumn,doublespacing,dvipsnames]{bmcart}%

\usepackage{amsthm,amsmath}
\usepackage[utf8]{inputenc} %
\usepackage{cite}

\usepackage{amsmath,amssymb,amsfonts,mathtools}

\DeclareMathAlphabet{\mathup}{OT1}{\familydefault}{m}{n}

\DeclareMathOperator{\Diag}{Diag}

\DeclareMathOperator{\vech}{vech}

\DeclareMathOperator{\expectationApprox}{\hat{\mathbb{E}}}
\DeclareMathOperator{\pearsonCorr}{\mathcal{R}}
\DeclareMathOperator{\cut}{cut}
\DeclareMathOperator{\vol}{vol}
\DeclareMathOperator{\ncut}{ncut}

\DeclareMathOperator{\entropy}{\mathcal{H}}

\newcommand*{\decoTran}{\mathup{T}}

\newcommand*{\decoConj}{*}

\newcommand*{\realNumbers}{\mathbb{R}}

\newcommand{\estimate}[1]{\hat{#1}}

\newcommand*{\vect}[1]{\mathbf{#1}}  %
\newcommand*{\mat} [1]{\mathbf{#1}}  %

\newcommand*{\given}{\big|}

\newcommand*{\idxTime}{t}  %
\newcommand*{\idxFrame}{k} %
\newcommand*{\numFrame}{K}
\newcommand*{\idxFreq}{n}  %

\newcommand*{\idxCh}{j}
\newcommand*{\idxChAlt}{j'}
\newcommand*{\numCh}{J}

\newcommand*{\idxFeat}{f}
\newcommand*{\numFeat}{F}

\newcommand*{\numTrial}{R}
\newcommand*{\numTrialSim}{\numTrial_{\text{sim}}}
\newcommand*{\numTrialRec}{\numTrial_{\text{rec}}}

\newcommand*{\sampleFreq}{f_{\mathup{s}}}
\newcommand*{\blockLength}{L} %
\newcommand*{\blockShift}{M} %

\newcommand*{\reverbTime}{T_{60}}

\newcommand*{\sigTargetBase}{s}
\newcommand*{\sigTarget}{\sigTargetBase}

\newcommand*{\roomImpulseResponseBase}{h}
\newcommand*{\roomImpulseResponse}{\roomImpulseResponseBase}
\newcommand*{\transferFunctionBase}{\MakeUppercase{\roomImpulseResponseBase}}
\newcommand*{\transferFunction}[1]{\transferFunctionBase_{#1}}

\newcommand*{\sigNoiseBase}{n}
\newcommand*{\sigNoise}{\sigNoiseBase}

\newcommand*{\sigBase}{x}
\newcommand*{\sigCh}[1]{\sigBase_{#1}}

\newcommand*{\sigBlockCh}[1]{\vect{\sigBase}_{#1}}

\newcommand*{\snr}{\mathup{SNR}}

\newcommand*{\mscBase}{\gamma}
\newcommand*{\msc}{\mscBase}

\newcommand*{\psdBase}{\Phi}
\newcommand*{\psd}[1]{\psdBase_{#1}}

\newcommand*{\featBase}{a}
\newcommand*{\featElem}[1]{\featBase_{#1}}

\newcommand*{\featVect}{\vect{\featBase}}

\newcommand*{\featVecInstant}{\featVect}
\newcommand*{\featVecInstantMean}{\overline{\featVect}}
\newcommand*{\featVecRecursionFactor}{\lambda}

\newcommand*{\featCorrBase}{b}
\newcommand*{\featCorrRawCh}[1]{\featCorrBase_{#1}}
\newcommand*{\featCorrRawMat}{\mat{\MakeUppercase{\featCorrBase}}}

\newcommand*{\featCorrRawMatInstant}{\mat{\MakeUppercase{\featBase}}}

\newcommand*{\featCorrCh}[1]{\tilde{\featCorrBase}_{#1}}
\newcommand*{\featCorrMat}{\tilde{\mat{\MakeUppercase{\featCorrBase}}}}

\newcommand*{\featChanMat}{\mat{C}}
\newcommand*{\featChanSingVal}{\sigma}

\newcommand*{\featChanLeftSingVec}{\vect{r}}
\newcommand*{\featChanRightSingVec}{\vect{t}}

\newcommand*{\featEnergyBase}{e}
\newcommand*{\featEnergyCh}[1]{\featEnergyBase_{#1}}

\newcommand*{\featEnergyMat}{\mat{\MakeUppercase{\featEnergyBase}}}

\newcommand*{\numStates}{Q}
\newcommand*{\KFcovarianceMatBase}{\mat{\Sigma}}
\newcommand*{\KFstateVec}{\boldsymbol{\mu}}
\newcommand*{\KFstateVecLatent}{\vect{z}}
\newcommand*{\KFobservationVec}{\boldsymbol{\xi}}

\newcommand*{\KFpriorStateCovarianceMat}{\KFcovarianceMatBase_{\mathup{s}}} 
\newcommand*{\KFprocessNoiseVec}{\vect{t}}
\newcommand*{\KFprocessNoiseCovarianceMat}{\KFcovarianceMatBase_{\mathup{t}}}
\newcommand*{\KFobservationNoiseVec}{\vect{o}}
\newcommand*{\KFobservationNoiseCovarianceMat}{\KFcovarianceMatBase_{\mathup{o}}}
\newcommand*{\KFgain}{\mat{K}}

\newcommand*{\KFprocessNoiseCovarianceMatScale}{\alpha_1}
\newcommand*{\KFobservationNoiseCovarianceMatScale}{\alpha_2}

\newcommand*{\simBase}{R}
\newcommand*{\simMat}{\mat{\simBase}}

\newcommand*{\graph}{\mathcal{G}}

\newcommand*{\graphVertices}{\mathcal{V}}

\newcommand*{\graphEdges}{\mathcal{E}}
\newcommand*{\graphEdgeWeight}{w}
\newcommand*{\matGraphConn}{\mat{W}}
\newcommand*{\matGraphLaplacian}{\mat{L}}
\newcommand*{\graphDegree}{d}
\newcommand*{\matGraphDegree}{\mat{D}}
\newcommand*{\fiedlerBase}{v}
\newcommand*{\vecFiedler}{\vect{\fiedlerBase}}

\newcommand*{\utilityBase}{u}
\newcommand*{\vecUtility}{\vect{\utilityBase}}

\newcommand*{\pccUtilityEntropy}{\rho}
\newcommand*{\vecSupplemental}{\boldsymbol{\beta}}

\newcommand*{\indicatorVec}{\vect{i}}

\newcommand*{\idxHistBin}{n_{\mathup{B}}}
\newcommand*{\numHistBin}{N_{\mathup{B}}}
\newcommand*{\histBinEdge}{e}

\providecommand*{\probabilityDensity}{p}

\newcommand*{\corrUtilityCoherence}{r}
\newcommand*{\vecMSC}{\boldsymbol{\mscBase}}
\newcommand*{\vecMSCZeroMean}{\widetilde{\vecMSC}}

\newcommand*{\setAllCh}{\mathcal{P}}
\newcommand*{\setSelCh}{\mathcal{S}}

\newcommand*{\weightFeatureBase}{\phi}
\newcommand*{\vecWeightFeature}{\boldsymbol{\weightFeatureBase}}
\newcommand*{\costWeightFeature}{\mathcal{C}}
\newcommand*{\lassoWeight}{\delta}

\newcommand*{\setTrainableParameters}{\Psi}

\newcommand*{\methodModelProp}{\texttt{model-KF}\xspace}
\newcommand*{\methodModelSmooth}{\texttt{model-smooth}\xspace}
\newcommand*{\methodModelEstimatornet}{\texttt{hybrid}\xspace}
\newcommand*{\methodModelMSC}{\texttt{baseline-MSC}\xspace}
\newcommand*{\methodModelCDR}{\texttt{baseline-CDR}\xspace}
\newcommand*{\methodMLSynth}{\texttt{ML-sim}\xspace}
\newcommand*{\methodMLTuned}{\texttt{ML-tuned}\xspace}
\newcommand*{\methodMLJoint}{\texttt{ML-joint}\xspace}

\definecolor{LMSred}{rgb}{0.80,0.20,0.20} 
\definecolor{LMSgray}{rgb}{0.60,0.60,0.60}
\definecolor{LMSGray}{rgb}{0.60,0.60,0.60}

\definecolor{LMSdarkgray}{rgb}{0.35,0.35,0.35}

\definecolor{LMSfootergray}{RGB}{139,139,139} %

\definecolor{LMSCyan}{rgb}{0.60,1,1}
\definecolor{LMSMagenta}{rgb}{0.965,0.294,1}
\definecolor{LMSYellow}{rgb}{0.95,0.95,0.0}

\definecolor{LMSOrange}{rgb}{1,0.5,0}

\definecolor{LMSlightblue}{rgb}{0.5,0.8,1}
\definecolor{LMSlightred}{rgb}{1,0.294,0.294}

\definecolor{LMSblue}{rgb}{0.216,0.255,1}

\definecolor{LMSgreen}{rgb}{0.15,0.7,0.15}
\definecolor{LMSlightgreen}{rgb}{0.35,0.90,0.35}

\usepackage{graphicx}

\usepackage{algorithm,algorithmic}

\usepackage{textcomp,xspace,siunitx,url}

\usepackage{enumitem}

\usepackage{subcaption,caption}

\usepackage{booktabs,multirow}

\usepackage[style=index,savewrites=false,nohypertypes={acronym},acronym,nomain,automake,nopostdot,nonumberlist,nogroupskip]{glossaries}
\makeglossaries

\newacronym{aec}{AEC}{Acoustic Echo Cancellation}
\newacronym{ann}{ANN}{Artificial Neural Network}
\newacronym{ap}{AP}{Access Point}
\newacronym{asn}{ASN}{Acoustic Sensor Network}
\newacronym{asr}{ASR}{Automatic Speech Recognition}
\newacronym{bss}{BSS}{Blind Source Separation}
\newacronym{cdr}{CDR}{Coherent-to-Diffuse power Ratio}
\newacronym{clt}{CLT}{Central Limit Theorem}
\newacronym[firstplural=Cross-Power Spectral Densities]{cpsd}{CPSD}{Cross-Power Spectral Density} 
\newacronym{dft}{DFT}{Discrete Fourier Transform}
\newacronym{doa}{DOA}{Direction of Arrival}
\newacronym{drr}{DRR}{Direct-to-Reverberation Ratio}
\newacronym{dsb}{DSB}{Delay-and-Sum Beamformer}
\newacronym{evd}{EVD}{Eigenvalue Decomposition}
\newacronym{erle}{ERLE}{Echo Return Loss Enhancement}
\newacronym{fft}{FFT}{Fast Fourier Transform}
\newacronym{fir}{FIR}{Finite Impulse Response}
\newacronym{gru}{GRU}{Gated Recurrent Unit}
\newacronym{gsc}{GSC}{Generalized Sidelobe Canceller}
\newacronym{ica}{ICA}{Independent Component Analysis}
\newacronym{iva}{IVA}{Independent Vector Analysis}
\newacronym{kf}{KF}{Kalman Filter}
\newacronym{lasso}{LASSO}{Least Absolute Shrinkage and Selection Operator}
\newacronym{lcmv}{LCMV}{Linearly Constrained Minimum Variance}
\newacronym{lmmse}{LMMSE}{Linear Minimum Mean Square Error}
\newacronym{ls}{LS}{Least Squares}
\newacronym{marvelo}{MARVELO}{Multicast-Aware Routing for Virtual network Embedding with Loops in Overlays}
\newacronym{mccc}{MCCC}{MultiChannel Correlation Coefficient}
\newacronym{ml}{ML}{Machine Learning}
\newacronym{mmse}{MMSE}{Minimum Mean Square Error}
\newacronym{msc}{MSC}{Magnitude-Squared Coherence}
\newacronym{mse}{MSE}{Mean Square Error}
\newacronym{music}{MUSIC}{MUltiple SIgnal Classification}
\newacronym{mvdr}{MVDR}{Minimum Variance Distortionless Response}
\newacronym{pcc}{PCC}{Pearson Correlation Coefficient}
\newacronym{pdf}{PDF}{Probability Density Function}
\newacronym{pi}{PI}{Power Iteration}
\newacronym{ppm}{PPM}{Parts Per Million}
\newacronym[firstplural=Power Spectral Densities]{psd}{PSD}{Power Spectral Density} 
\newacronym{roi}{RoI}{Region of Interest}
\newacronym{rir}{RIR}{Room Impulse Response}
\newacronym{sdr}{SDR}{Signal-to-Distortion Ratio}
\newacronym{sinr}{SINR}{Signal-to-Interference-plus-Noise Ratio}
\newacronym{sir}{SIR}{Signal-to-Interference Ratio}
\newacronym{snr}{SNR}{Signal-to-Noise Ratio}
\newacronym[plural=SOIs,firstplural=Sources of Interest]{soi}{SOI}{Source of Interest}
\newacronym{spp}{SPP}{Speech Presence Probability}
\newacronym{stft}{STFT}{Short-Time Fourier Transform}
\newacronym{svd}{SVD}{Singular Value Decomposition}
\newacronym{vne}{VNE}{Virtual Network Embedding}
\newacronym{wasn}{WASN}{Wireless Acoustic Sensor Network}
\newacronym{wer}{WER}{Word Error Rate}

\usepackage{tikz}
\usetikzlibrary{arrows.meta,backgrounds,calc,decorations.markings,fit,patterns,positioning,shapes,trees}
\tikzset{>=stealth} %
\usepackage{LMSnodeshapes} %

\usetikzlibrary{external}
\usepackage{pgfplots}
\pgfplotsset{compat=1.3}
\usepgfplotslibrary{fillbetween}

\usepackage{cleveref}
\Crefname{equation}{}{}
\crefname{equation}{}{}
\Crefname{figure}{Fig.}{Figs.}
\crefname{figure}{Fig.}{Figs.}
\Crefname{section}{Section}{Sections}
\crefname{section}{Section}{Sections}
\Crefname{table}{Table}{Tables}
\crefname{table}{Table}{Tables}

\newcommand{\eg}{e.\,g.\xspace}
\newcommand{\ie}{i.\,e.\xspace}
\newcommand{\wrt}{w.\,r.\,t.\xspace}

\startlocaldefs
\endlocaldefs

\begin{document}

\begin{frontmatter}

\begin{fmbox}
\dochead{Research}

\title{Microphone Utility Estimation in Acoustic Sensor Networks using Single-Channel Signal Features}

\author[
  addressref={aff1},                   %
  corref={aff1},                       %
  email={michael.guenther@fau.de}   %
]{\inits{M.G.}\fnm{Michael} \snm{Günther}}
\author[
  addressref={aff1},
  email={andreas.brendel@fau.de}
]{\inits{A.B.}\fnm{Andreas} \snm{Brendel}}
\author[
  addressref={aff1},
  email={walter.kellermann@fau.de}
]{\inits{W.K.}\fnm{Walter} \snm{Kellermann}}

\address[id=aff1]{%
  \orgdiv{Chair of Multimedia Communications and Signal Processing},             %
  \orgname{Friedrich-Alexander-Universität},  %
  \city{Erlangen},                            %
  \cny{DE}                                    %
}

\begin{abstractbox}

\begin{abstract} %
In multichannel signal processing with distributed sensors, choosing the optimal subset of observed sensor signals to be exploited is crucial in order to maximize algorithmic performance and reduce computational load, ideally both at the same time.
In the acoustic domain, signal cross-correlation is a natural choice to quantify the usefulness of microphone signals, \ie, microphone utility, for array processing, but its estimation requires that the uncoded signals are synchronized and transmitted between nodes.
In resource-constrained environments like acoustic sensor networks, low data transmission rates often make transmission of all observed signals to the centralized location infeasible, thus discouraging direct estimation of signal cross-correlation.
Instead, we employ characteristic features of the recorded signals to estimate the usefulness of individual microphone signals.
In this contribution, we provide a comprehensive analysis of model-based microphone utility estimation approaches that use signal features and, as an alternative, also propose machine learning-based estimation methods that identify optimal sensor signal utility features.
The performance of both approaches is validated experimentally using both simulated and recorded acoustic data, comprising a variety of realistic and practically relevant acoustic scenarios including moving and static sources.
\end{abstract}

\begin{keyword}
\kwd{channel selection}
\kwd{graph partitioning}
\kwd{microphone utility}
\kwd{acoustic sensor network}
\end{keyword}

\end{abstractbox}
\end{fmbox}%

\end{frontmatter}

\section{Introduction}
An \gls{asn} comprises multiple spatially distributed microphones that typically communicate wirelessly.
Capturing different perspectives of the acoustic scene, the signals recorded by these distributed microphones encode spatial information exploitable by multichannel signal processing algorithms.
These algorithms accomplish crucial tasks \cite{bertrand_applications_2011} like acoustic source localization \cite{wang_voice_1997,cheng_survey_2012,brendel_distributed_2019} and tracking \cite{kaplan_maximum_2001,evers_locata_2020}, extraction and enhancement of an acoustic \gls{soi} \cite{brandstein_microphone_2001,bertrand_distributed_2009,markovich-golan_optimal_2015}, hands-free communication \cite{gay_acoustic_2012}, acoustic monitoring \cite{oliveira_wireless_2011,goetze_acoustic_2012}, as well as scene classification and acoustic event detection \cite{mesaros_dcase_2017}.
As the microphones in \glspl{asn} often have no common sampling clock, their signals must be synchronized before joint processing.

The performance of these signal processing algorithms is affected by many factors including the proximity of the microphones to desired and undesired acoustic sources, reverberation, additive noise, orientation and occlusion of microphones, among others.
As a result, the signals obtained from different microphones are generally not equally useful for the above-mentioned tasks, potentially even detrimental in extreme cases if inappropriate importance is assigned to them.
To ensure optimal algorithmic performance at minimum transmission cost and computational cost, a diligent selection which of the observed microphone signals to process and which to discard is crucial in order to avoid unnecessary signal transmission or synchronization efforts.
Unsurprisingly, this task has received considerable attention in the literature: the selection of a single best channel for \gls{asr} based on signal features has been explored in \cite{wolf_towards_2009}.
A utility measure specifically for \gls{mmse} signal extraction has been proposed in \cite{bertrand_efficient_2010,szurley_energy_2012}, followed by a distributed version in \cite{szurley_greedy_2014}.
Microphone subset selection to minimize communication cost with upper-bounded output noise power has been investigated for both \gls{mvdr} \cite{zhang_microphone_2018} and \gls{lcmv} \cite{zhang_sensor_2021} beamforming.
More recently, joint microphone subset selection and speech enhancement using deep learning was proposed in \cite{casebeer_communication-cost_2021}.
However, these methods either neglect the limitations of the underlying \gls{asn} regarding communication cost or are tailored to a specific application.
In the following, we present a different approach that overcomes both drawbacks, \ie, requires little transmission data rate and is applicable to a broad class of signal processing applications.

Many multichannel algorithms, \eg, for signal enhancement or localization, \cite{brandstein_microphone_2001,benesty_microphone_2008} assume coherent, \ie, linearly related, input signals and exploit the spatial information captured by the inter-channel phase differences.
Thus, the cross-correlation of microphone signal pairs and measures derived from it, in particular the spatial coherence and the \gls{msc}, are intuitive measures for quantifying the usefulness of observed microphone signals and have been used in literature for that purpose, \eg, in \cite{kumatani_channel_2011}.
For synchronized microphones with sufficient transmission data rate, direct estimation of the inter-channel coherence from the observed uncoded microphone signals to rate their utility is straightforward.
However, in \glspl{asn}, this approach is often precluded by a limited transmission data rate, \eg, of current wireless networks \cite{noauthor_ieee_2016}, especially when the number of microphones is large.
This issue is further compounded if the available data rate must be shared with other, possibly non-audio, applications, like video streaming in smart home environments.
Furthermore, the microphone signals in \glspl{asn} generally do not share a common sampling clock \cite{chinaev_online_2021}, thus requiring potentially costly synchronization prior to estimation of the coherence.
Clearly, this disqualifies direct estimation of the signal cross-correlation as a viable approach for estimating microphone utility in \glspl{asn} when promising candidate microphone signals for synchronization and subsequent joint processing should be identified.

Therefore, to address these unique challenges of \glspl{asn}, we employ a compressed signal representation in the form of single-channel signal feature sequences, which are extracted from temporal blocks of the microphone signals, to reduce the amount of data to be transmitted.
The employed features must be characteristic for the microphone signals, \ie, to allow for (at least approximate) reconstruction of the inter-channel \gls{msc}.

In this contribution, we consider acoustic scenarios often encountered in smart home applications comprising a single \gls{soi} captured by multiple distributed microphones in an acoustic enclosure, as depicted in \cref{fig:introduction-scenario}.
After estimating the usefulness of the recorded microphone signals, a subset thereof is selected and transmitted to the central wireless \gls{ap} for subsequent coherent multichannel signal processing.
Although \cref{fig:introduction-scenario} shows an exemplary scenario with a wireless network with a central \gls{ap}, this is not constraining the scope of the paper.
For the considerations in this paper, the network may also be wired and the \gls{ap} may be replaced by a network node acting as a local center implementing the multichannel signal processing algorithm.
Instead of a specific signal processing algorithm, we consider a broad class of algorithms that relies on coherent input signals and does not require signals unrelated to the \gls{soi}, \eg, as noise references.
In addition, we do not consider application-specific cost functions or performance metrics, \eg, \gls{sdr} and \gls{erle}, such that the proposed utility estimation scheme is appropriate for many subsequent multichannel signal processing applications.
Thus, the proposed generic system takes the form depicted in \cref{fig:system-overview} comprising two main subsystems: a feature extraction system, a copy of which runs for each microphone signal on the associated network node, and a utility estimation system running on the \gls{ap}.
In the feature extraction stage, characteristic signal features are extracted from the observed microphone signals independently from each other, \ie, no cross-channel features are employed.
The feature sequences obtained from each microphone are then transmitted to the central \gls{ap}, which estimates the individual microphones' utility values.
A joint approach additionally considering network transmission cost was proposed in \cite{gunther_network-aware_2021}, while the efficacy of the proposed utility estimates for robust source localization and spatial filtering was demonstrated in \cite{gunther_microphone_2021} and \cite{afifi_reinforcement_2021}, respectively.

In the remainder of this article, we review and provide more detailed descriptions of the model-based realizations of the two subsystems proposed in \cite{gunther_network-aware_2021,gunther_online_2021} in \cref{sec:model-feature-extraction,sec:model-utility-estimation} by explicitly stating and discussing the model assumptions of the \gls{kf}.
By formulating microphone selection as a graph bi-partitioning problem, the Fiedler vector is interpreted as an optimum soft assignment of the microphones to the two groups of most and least useful microphones, further justifying its use as a utility measure.
In \cref{sec:feature-importance}, we provide new results on the suitability of established signal features for recovering inter-channel \gls{msc}.
To this end, the feature selection task is formulated as a \gls{lasso} regression problem which is then solved numerically to obtain an optimal set of signal features.
In \cref{sec:learning-based}, we propose novel \gls{ml}-based realizations for both subsystems whose combination can be learned in an end-to-end fashion, which constitutes a major contribution of this work.
In \cref{sec:experiments}, the efficacy of the proposed scheme and its individual components is validated.
Different algorithmic variants, \ie, purely model-based, purely \gls{ml}-based, and hybrid realizations of the proposed system, are investigated.
To this end, comprehensive experiments for both synthesized and recorded data from realistic scenarios are conducted, including different reverberation times, additive noise and obstruction of sensors, different microphone arrangements, as well as static and moving \glspl{soi}.

\section{Notation and Signal Model}
In this article, scalar quantities are denoted by slanted non-bold symbols $x$, while vectors and matrices are denoted by bold-face lowercase $\vect{x}$ and uppercase symbols $\mat{X}$, respectively.
Furthermore, $[\vect{x}]_{\idxCh}$ denotes the $\idxCh$-th element of vector $\vect{x}$, and $[\mat{X}]_{\idxCh,\idxChAlt}$ denotes the ($\idxCh$,$\idxChAlt$)-th element of matrix $\mat{X}$.
The $J$-dimensional all-zeros and all-ones vectors are denoted by $\vect{0}_{J}$ and $\vect{1}_{J}$, respectively, the $J \times J$ identity matrix is denoted by $\mat{I}_{J}$, and the operator $\Diag(\cdot)$ embeds the elements of its argument on the main diagonal of a square matrix.
The \gls{pcc} of two $J$-element vectors $\vect{x}$, $\vect{y}$ is defined as
\begin{equation}\hspace*{-5pt}
\pearsonCorr(\vect{x},\vect{y}) = \frac%
{\sum_{j=1}^{J} ([\vect{x}]_{j} - \overline{x}) ([\vect{y}]_{j} - \overline{y})} %
{\sqrt{ \sum_{j=1}^{J} ([\vect{x}]_{j} - \overline{x})^2} \sqrt{\sum_{j=1}^{J} ([\vect{y}]_{j} - \overline{y})^2 }}
\label{eq:pcc-definition-compact}
\end{equation}
with means $\overline{x} = \frac{1}{J} \sum_{j=1}^{J}[\vect{x}]_{j}$ and $\overline{y} = \frac{1}{J} \sum_{j=1}^{J}[\vect{y}]_{j}$.

In the following, let $\idxTime$ denote the discrete-time sample index and let $\idxFeat \in \{1,\ldots,\numFeat\}$ denote the feature index where $\numFeat$ is the number of extracted features per channel.
Recalling \cref{fig:introduction-scenario}, we consider an acoustic scenario comprising a single coherent \gls{soi} recorded by $\numCh$ microphones, each of which represents a separate node in the \gls{asn}.
The signal captured by the microphone indexed by $\idxCh \in \setAllCh = \{1,\ldots,\numCh\}$ is
\begin{equation}
\sigCh{\idxCh}[\idxTime] = \sigTarget[\idxTime] * \roomImpulseResponse_{\idxCh}[\idxTime] + \sigNoise_{\idxCh}[\idxTime],
\label{eq:signal-model}
\end{equation}
where $\sigTarget[\idxTime]$ is the dry \gls{soi} signal, $\roomImpulseResponse_{\idxCh}[\idxTime]$ is the acoustic impulse response from the \gls{soi} to the $\idxCh$-th microphone, and $*$ denotes linear convolution.
Note that the \gls{soi} is not necessarily static, \ie, the acoustic impulse responses $\roomImpulseResponse_{\idxCh}[\idxTime]$ in \cref{eq:signal-model} are considered time-invariant only for short observation intervals, but may change from one interval to the next as the \gls{soi} moves.
The fully coherent spatial images of the \gls{soi} $\sigTarget[\idxTime] * \roomImpulseResponse_{\idxCh}[\idxTime]$ are superimposed by a spatially diffuse or incoherent noise field, such that the mutual coherence between the noise components $\sigNoise_{\idxCh}[\idxTime], \;\forall\idxCh\in\setAllCh$ is negligibly small.
Thus, observed correlation between two microphone signals $\sigCh{\idxCh}[\idxTime]$, $\sigCh{\idxChAlt}[\idxTime]$ is predominantly caused by the common \gls{soi} signal.
Although competing point-like sources are not explicitly modeled in \cref{eq:signal-model}, the model can still yield useful and intuitively desired results in this case.
For example, consider an \gls{asn} spanning two rooms within an apartment connected by, \eg, open doors.
With the microphones in each room predominantly capturing the corresponding \gls{soi}, the realizations of the proposed system in \cref{sec:model-based,sec:learning-based} can still facilitate a distinction of microphones \wrt the dominant \gls{soi}.
Nevertheless, the signal model \cref{eq:signal-model} represents a first step towards developing methods for more general acoustic scenarios.

As the proposed utility estimation relies only on the correlation of features sequences computed from signal frames, only a coarse synchronization of the signal frames between different sensors has to be assured, such that the proposed scheme is practically relevant.
However, to compute the oracle \gls{msc} between the sensors for evaluation of the method, we assume that the sensors are perfectly synchronized.
The signals are partitioned into blocks indexed by $\idxFrame\in\{1,\ldots,\numFrame\}$ with a length of $\blockLength$ samples and a shift of $\blockShift$ between successive blocks, \eg, for the $\idxCh$-th microphone signal $\sigCh{\idxCh}[\idxTime]$,
\begin{equation}
\sigBlockCh{\idxCh}[\idxFrame] = \begin{bmatrix} \sigCh{\idxCh}[\idxFrame\blockShift], & \ldots, & \sigCh{\idxCh}[\idxFrame\blockShift+\blockLength-1]
\label{eq:signal-block}
\end{bmatrix}^\decoTran
\in \realNumbers^{\blockLength}.
\end{equation}
As a broadband ground-truth utility measure for the $\idxCh$-th microphone, the frequency-averaged narrowband \gls{msc}
\begin{equation}
\msc_{\idxCh}[\idxFrame] = 
\frac{1}{\blockLength} \sum_{\idxFreq=1}^{\blockLength} %
\left|
\frac{\psd{\sigTarget,\sigCh{\idxCh}}[\idxFrame,\idxFreq]} %
    {\sqrt{\psd{\sigTarget,\sigTarget}[\idxFrame,\idxFreq] \cdot \psd{\sigCh{\idxCh},\sigCh{\idxCh}}[\idxFrame,\idxFreq]}}
\right|^2
\label{eq:MSC-source-microphone}
\end{equation}
between the (latent) source signal $\sigTarget[\idxTime]$ and the $\idxCh$-th microphone signal $\sigCh{\idxCh}[\idxTime]$ is used.
Therein, $\idxFreq\in\{1,\ldots,\blockLength\}$ denotes the discrete frequency bin index, the \gls{dft} length is chosen identical to the block length $\blockLength$, and $\psd{\sigTarget,\sigCh{\idxCh}}[\idxFrame,\idxFreq]$, $\psd{\sigTarget,\sigTarget}[\idxFrame,\idxFreq]$ and $\psd{\sigCh{\idxCh},\sigCh{\idxCh}}[\idxFrame,\idxFreq]$ are short-time estimates of the cross-\gls{psd} and auto-\glspl{psd} of $\sigTarget[\idxTime]$ and $\sigCh{\idxCh}[\idxFrame]$, respectively.
Under the assumption that the \gls{soi} signal $\sigTarget[\idxTime]$ and noise signals $\sigNoise_{\idxCh}[\idxTime]$ are mutually uncorrelated, combined with \cref{eq:signal-model}, the relations
\begin{align}
\psd{\sigTarget,\sigCh{\idxCh}}[\idxFrame,\idxFreq] &= \transferFunction{\idxCh}^{\decoConj}[\idxFrame,\idxFreq] \, \psd{\sigTarget,\sigTarget}[\idxFrame,\idxFreq], \\
\psd{\sigCh{\idxCh},\sigCh{\idxCh}}[\idxFrame,\idxFreq] &= \left|\transferFunction{\idxCh}[\idxFrame,\idxFreq]\right|^{2} \, \psd{\sigTarget,\sigTarget}[\idxFrame,\idxFreq] + \psd{\sigNoise_{\idxCh},\sigNoise_{\idxCh}}[\idxFrame,\idxFreq]
\end{align}
hold, such that the summands in \cref{eq:MSC-source-microphone} simplify to
\begin{equation}
\left|
\frac{\psd{\sigTarget,\sigCh{\idxCh}}[\idxFrame,\idxFreq]} %
    {\sqrt{\psd{\sigTarget,\sigTarget}[\idxFrame,\idxFreq] \cdot \psd{\sigCh{\idxCh},\sigCh{\idxCh}}[\idxFrame,\idxFreq]}}
\right|^2
=
\frac{\snr_{\idxCh}[\idxFrame,\idxFreq]}{1 + \snr_{\idxCh}[\idxFrame,\idxFreq]},
\label{eq:MSC-SNR-relation}
\end{equation}
with the channel-wise \gls{snr}
\begin{equation}
\snr_{\idxCh}[\idxFrame,\idxFreq] = \frac{\left|\transferFunction{\idxCh}[\idxFrame,\idxFreq]\right|^{2} \,\psd{\sigTarget,\sigTarget}[\idxFrame,\idxFreq]}{\psd{\sigNoise_{\idxCh},\sigNoise_{\idxCh}}[\idxFrame,\idxFreq]}.
\label{eq:snr-channel}
\end{equation}
Clearly, the \gls{msc} is a function of the \gls{snr}, with extremal values $\msc_{\idxCh}[\idxFrame,\idxFreq]=0$ for $\snr_{\idxCh}[\idxFrame,\idxFreq]=0$, and $\msc_{\idxCh}[\idxFrame,\idxFreq]\rightarrow 1$ for $\snr_{\idxCh}[\idxFrame,\idxFreq]\rightarrow\infty$.
The frequency-averaged \gls{msc} values of all $\numCh$ microphones are collected in the vector
\begin{equation}
\vecMSC[\idxFrame] = \begin{bmatrix} \msc_{1}[\idxFrame], & \ldots, & \msc_{\numCh}[\idxFrame] \end{bmatrix}^{\decoTran} \in [0,1]^{\numCh}.
\label{eq:MSC-vector}
\end{equation}

\section{Model-based Utility Estimation using Spectral Graph Partitioning}
\label{sec:model-based}
We first review the model-based realizations of \cite{gunther_network-aware_2021,gunther_online_2021} in \cref{sec:model-feature-extraction,sec:model-utility-estimation}.
Although there is no strictly analytical relation between the extracted feature values and the utility values, the approach is based on the notion that the \glspl{pcc} of the different extracted feature sequences all reflect the same pair-wise similarity of the underlying microphone signals.
Due to this model assumption, and to differentiate it from the \gls{ml}-based approach in \cref{sec:learning-based}, which requires training data to determine the model parameters, this approach is termed model-based.
Advancing the previous heuristic feature selection\cite{gunther_single-channel_2019}, we formulate the feature selection task as a \gls{lasso} regression problem in \cref{sec:feature-importance} to optimize the trade-off between accuracy and number of features to be transmitted.
Solving this problem numerically yields an optimal selection of features for a set of representative acoustic scenarios.

\subsection{Node-wise Feature Extraction}
\label{sec:model-feature-extraction}
There is a wide variety of potential signal features \cite{peeters2004large,virtanen_computational_2018} to describe acoustic signals.
Since acoustic scenarios are typically not static in practice due to, \eg, moving acoustic sources or obstructions, the usefulness of microphones is equally time-variant.
Hence, within the comprehensive feature taxonomy in \cite{peeters2004large}, we focus on features extracted from short observation intervals to characterize single-channel signals.
The features may be computed in the time domain based on the digital signal waveform, or in the frequency domain based on the magnitude spectrum of the signals.
As a result, we consider the following block-wise features:
\begin{itemize}
\item envelope of waveform
\item zero-crossing rate
\item statistical moments (centroid, standard deviation, skewness, kurtosis) of the signal waveform
\item entropy of waveform
\item statistical moments (centroid, standard deviation, skewness, kurtosis) of the magnitude spectrum
\item spectral shape features (slope, power flatness, amplitude flatness, roll-off)
\item temporal variation of magnitude spectra (spectral flux, spectral flux of normalized magnitude spectra, spectral variation)
\end{itemize}
In \cite{gunther_single-channel_2019}, it was experimentally shown that three features (temporal skewness, temporal kurtosis, spectral flux) are suitable to recover the structure of the spatial \gls{msc} matrix of a set of microphone signals.
However, the features were selected heuristically based on the visual similarity of the corresponding feature covariance matrices and the ground-truth \gls{msc} matrix.
Therefore, a more rigorous discussion of the importance of specific signal features is provided in \cref{sec:feature-importance}.

Generally, a single feature sequence, \ie, a sequence of feature values over several time frames, is insufficient to characterize the signals, since the extraction of each signal feature can at best maintain information about the original signal \cite{cover_elements_2006}, but typically incurs a loss of information.
When multiple sufficiently different features are used, they capture different parts of the information contained in the signals, such that they complement each other in describing the original signals.
Thus, jointly processing such different features allows for a more accurate characterization of the signals compared to a single feature.

To this end, given a signal block $\sigBlockCh{\idxCh}[\idxFrame]$, let $\featBase_{\idxCh}^{(\idxFeat)}[\idxFrame]$ denote the observed value of the signal feature $\idxFeat\in\{1,\ldots,\numFeat\}$ for said signal block.
Collecting the feature values of different channels for time frame $\idxFrame$ yields the instantaneous feature vector
\begin{align}
\featVecInstant^{(\idxFeat)}[\idxFrame] &= \begin{bmatrix} \featElem{1}^{(\idxFeat)}[\idxFrame], & \ldots, & \featElem{\numCh}^{(\idxFeat)}[\idxFrame] \end{bmatrix}^{\decoTran} \in \realNumbers^{\numCh}.
\label{eq:feature-vector-instant}
\end{align}
Unlike the signal waveforms, which require precise synchronization of the sampling clocks for joint processing, the feature values of different microphones are much less susceptible to asynchronous sampling.
With only a single feature value every $\blockShift$ signal samples, sampling rate offsets on the order of tens of \gls{ppm} barely affect the extracted feature sequences.
Hence, periodical coarse synchronization of the signal block boundaries is sufficient to avoid excessive drift of the observation windows in different network nodes, allowing synchronization to occur less frequently and with lower accuracy requirements.

\subsection{Utility Estimation}
\label{sec:model-utility-estimation}
In this section, we review the utility estimation scheme based on correlation of feature sequences originally proposed in \cite{gunther_single-channel_2019,gunther_network-aware_2021,gunther_online_2021}, and show its relation to established Graph Bisection techniques.
The model-based utility estimation comprises three steps, which are outlined in the following subsections:
\begin{enumerate}
\item estimate the cross-channel \glspl{pcc} of the feature sequences separately for each feature
\item fuse information contained in the \glspl{pcc} from different features %
\item estimate each microphone's utility from the fused information by means of Spectral Graph Partitioning
\end{enumerate}
For clarity, a visual guide of these steps and the involved matrices and vectors is provided in \cref{fig:overview-model-utility-estimation}.

\textit{Step 1) Feature Correlation Coefficients:}
For computing the \glspl{pcc}, first the cross-channel covariance matrices
\begin{align}
\featCorrRawMat^{(\idxFeat)}[\idxFrame] %
&= \begin{bmatrix} %
    \featCorrRawCh{1,1}^{(\idxFeat)}[\idxFrame] & \cdots & \featCorrRawCh{1,\numCh}^{(\idxFeat)}[\idxFrame] \\
    \vdots & & \vdots \\
    \featCorrRawCh{\numCh,1}^{(\idxFeat)}[\idxFrame] & \cdots & \featCorrRawCh{\numCh,\numCh}^{(\idxFeat)}[\idxFrame]
\end{bmatrix}
= \expectationApprox\left( \featCorrRawMatInstant^{(\idxFeat)}[\idxFrame] \right)
\label{eq:feature-covariance-matrix}
\end{align}
are estimated for each feature $\idxFeat \in \{1,\ldots,\numFeat\}$ separately.
Therein, $\expectationApprox$ denotes an approximate statistical expectation operator whose practical realization we discuss below.
Furthermore, the matrix
\begin{align}
\featCorrRawMatInstant^{(\idxFeat)}[\idxFrame] &= \left(\featVecInstant^{(\idxFeat)}[\idxFrame] - \featVecInstantMean^{(\idxFeat)}[\idxFrame]\right)   \left(\featVecInstant^{(\idxFeat)}[\idxFrame] - \featVecInstantMean^{(\idxFeat)}[\idxFrame] \right)^{\decoTran}%
\label{eq:feature-covariance-outerproduct}
\end{align}
is the outer product of the instantaneous observed feature vector $\featVecInstant^{(\idxFeat)}[\idxFrame]$ after subtracting its recursive temporal average
\begin{equation}
\featVecInstantMean^{(\idxFeat)}[\idxFrame+1] = \featVecRecursionFactor \featVecInstantMean^{(\idxFeat)}[\idxFrame] + (1-\featVecRecursionFactor) \featVecInstant^{(\idxFeat)}[\idxFrame+1],
\label{eq:feature-vec-recursive}
\end{equation}
controlled by the recursive averaging factor $\featVecRecursionFactor \in [0,1]$ with initial value $\featVecInstantMean^{(\idxFeat)}[0] = \vect{0}_{\numCh}$.

Note that the estimated $\featCorrRawMat^{(\idxFeat)}[\idxFrame]$ is generally time-variant to account for the aforementioned \gls{soi} movement, and thus online estimation is preferred over batch estimation.
In order to track this temporal variability, we use a separate \gls{kf} \cite{kalman_new_1960} for each feature $\idxFeat$.
Let the latent state vector at time frame $\idxFrame$ be denoted by $\KFstateVecLatent^{(\idxFeat)}[\idxFrame]$.
Its mean vector $\KFstateVec^{(\idxFeat)}[\idxFrame]$ captures the covariance matrix $\featCorrRawMat^{(\idxFeat)}[\idxFrame]$ to be estimated and the instantaneous observation vector $\KFobservationVec^{(\idxFeat)}[\idxFrame]$ captures the matrix $\featCorrRawMatInstant^{(\idxFeat)}[\idxFrame]$.
Since both $\featCorrRawMat^{(\idxFeat)}[\idxFrame]$ and $\featCorrRawMatInstant^{(\idxFeat)}[\idxFrame]$ are symmetric, it is sufficient to only consider their non-redundant elements. 
We choose their diagonal elements and lower triangular elements, such that the dimensionality of the state vector $\KFstateVecLatent^{(\idxFeat)}[\idxFrame]$, the mean vector $\KFstateVec^{(\idxFeat)}[\idxFrame]$ and the observation vector $\KFobservationVec^{(\idxFeat)}[\idxFrame]$ can be chosen to be only $\numStates = \frac{\numCh (\numCh-1)}{2}$ instead of $\numCh^2$ while still precisely modeling the full matrices.
This can be expressed compactly using the \emph{half-vectorization} operator $\vech$ \cite{henderson_vec_1979}, \ie,
\begin{align}
\KFstateVec^{(\idxFeat)}[\idxFrame] %
&= \vech \left( \featCorrRawMat^{(\idxFeat)}[\idxFrame] \right) \in \realNumbers^{\numStates},
\label{eq:KF-state-vector} \\
\KFobservationVec^{(\idxFeat)}[\idxFrame] %
&= \vech\left( \featCorrRawMatInstant^{(\idxFeat)}[\idxFrame] \right) \in \realNumbers^{\numStates}.
\label{eq:KF-observation-vector}
\end{align}
The state-transition and model equations of the \gls{kf} are
\begin{align}
\KFstateVecLatent^{(\idxFeat)}[\idxFrame+1] &= \KFstateVecLatent^{(\idxFeat)}[\idxFrame] + \KFprocessNoiseVec^{(\idxFeat)}[\idxFrame],
\label{eq:KF-state-model} \\
\KFobservationVec^{(\idxFeat)}[\idxFrame] &= \KFstateVecLatent^{(\idxFeat)}[\idxFrame] + \KFobservationNoiseVec^{(\idxFeat)}[\idxFrame],
\label{eq:KF-observation-model}
\end{align}
where $\KFprocessNoiseVec^{(\idxFeat)}[\idxFrame]$ and $\KFobservationNoiseVec^{(\idxFeat)}[\idxFrame]$ denote the state-transition and observation noise vectors, respectively.
In other words, the most probable state transition is that the utility, and hence the feature covariance, stays the same.
However, if the state does change, it has no predictable preference direction.
Similarly simple motion models are effectively used in acoustic echo cancellation \cite{enzner_acoustic_2014} and dereverberation \cite{schwartz_online_2015}.
With \cref{eq:KF-state-model,eq:KF-observation-model}, the \gls{kf} simplifies to temporal smoothing, albeit with a time-variant smoothing constant.
Compared to fixed averaging constants, this allows placing higher confidence in observations with high signal energy, \ie, likely \gls{soi} activity.

Assuming a normally distributed latent state vector $\KFstateVecLatent^{(\idxFeat)}[\idxFrame]$ like in \cite{gunther_online_2021} for simplicity and mathematical tractability leads to the prior distribution
\begin{equation}
p\left(\KFstateVecLatent^{(\idxFeat)}[\idxFrame]\right) = \mathcal{N}\left(\KFstateVecLatent^{(\idxFeat)}[\idxFrame] \;\given\; \KFstateVec^{(\idxFeat)}[\idxFrame], \KFpriorStateCovarianceMat^{(\idxFeat)}[\idxFrame]\right)
\end{equation}
with the aforementioned mean vector $\KFstateVec^{(\idxFeat)}[\idxFrame]$ and covariance matrix $\KFpriorStateCovarianceMat^{(\idxFeat)}[\idxFrame] \in \realNumbers^{\numStates \times \numStates}$.
Since the trend of $\KFstateVecLatent^{(\idxFrame)}[\idxFrame]$ is neither known nor easily modeled, we assume a zero-mean Gaussian random walk with transition distribution
\begin{align}
p\left(\KFstateVecLatent^{(\idxFeat)}[\idxFrame+1] \;\given\; \KFstateVecLatent^{(\idxFeat)}[\idxFrame]\right) &= \mathcal{N}\left( \KFstateVecLatent^{(\idxFeat)}[\idxFrame+1] \;\given\; \KFstateVecLatent^{(\idxFeat)}[\idxFrame], \KFprocessNoiseCovarianceMat \right)
\label{eq:KF-transition-distribution}
\end{align}
as it is the least informative model but, due to the \gls{clt} \cite{papoulis_probability_2001}, fits well for natural processes where changes in the latent state are often the result of many independent influences.
In order to remain agnostic to the source-microphone arrangement in different scenarios, the time-invariant and feature-independent process noise covariance matrix %
is chosen as a scaled identity matrix
\begin{equation}
\KFprocessNoiseCovarianceMat = \KFprocessNoiseCovarianceMatScale \mat{I}_{\numStates} \in \realNumbers^{\numStates\times\numStates},
\label{eq:KF-state-noise}
\end{equation}
where $\KFprocessNoiseCovarianceMatScale \in \realNumbers^{+}$ is a positive tunable parameter.
Intuitively, two closely-spaced microphones produce similar feature sequences and thus the way their estimated \glspl{pcc} \wrt a third microphone change over time will be correlated.
While these scenario-specific correlations could in principle be exploited for more accurate estimation by tailoring $\KFprocessNoiseCovarianceMat$ to the scenario, doing so would harm the generalization of the transition model to other scenarios and furthermore requires the acquisition of sufficient data to estimate an optimal $\KFprocessNoiseCovarianceMat$.
Therefore, to avoid biasing the random walk process, we choose not to model these correlations, \ie, keep $\KFprocessNoiseCovarianceMat$ diagonal.

Choosing, for simplicity, the least informative emission model for the observations $\KFobservationVec^{(\idxFeat)}[\idxFrame]$ for simplicity as well yields the multivariate Gaussian emission distribution
\begin{align}
p\left(\KFobservationVec^{(\idxFeat)}[\idxFrame] \;\given\; \KFstateVecLatent^{(\idxFeat)}[\idxFrame]\right) &= \mathcal{N}\left( \KFobservationVec^{(\idxFeat)}[\idxFrame] \;\given\; \KFstateVecLatent^{(\idxFeat)}[\idxFrame], \KFobservationNoiseCovarianceMat[\idxFrame] \right)
\label{eq:KF-emission-distribution}
\end{align}
with the observation noise covariance matrix %
\begin{equation}
\KFobservationNoiseCovarianceMat[\idxFrame] = \KFobservationNoiseCovarianceMatScale \left( \Diag\left( \strut \vech\left(  \featEnergyMat[\idxFrame] \right) \right) \right)^{-1} \in\realNumbers^{\numStates\times\numStates}.
\label{eq:KF-observation-noise}
\end{equation}
Therein, $\KFobservationNoiseCovarianceMatScale \in \realNumbers^{+}$ is a positive tunable parameter and the matrix $\featEnergyMat[\idxFrame] \in \realNumbers^{\numCh\times\numCh}$ contains the geometric means of signal frame energies $\featEnergyCh{\idxCh}[\idxFrame] = \lVert \sigBlockCh{\idxCh}[\idxFrame] \rVert_{2}^{2}$ (see \cref{eq:signal-block}) reflecting the signal variances for each microphone pair %
\begin{equation}
\left[\strut\featEnergyMat[\idxFrame]\right]_{\idxCh\idxChAlt} = \sqrt{\featEnergyCh{\idxCh}[\idxFrame] \cdot \featEnergyCh{\idxChAlt}[\idxFrame]} + \epsilon, %
\quad\forall \idxCh, \idxChAlt \in \setAllCh.
\label{eq:energy-mean-matrix}
\end{equation}
The small positive constant $\epsilon$ ensures invertibility of $\KFobservationNoiseCovarianceMat[\idxFrame]$ in \cref{eq:KF-observation-noise} during speech absence periods.
This choice is motivated by the notion that the observed feature values are better at characterizing the microphone signals during time frames with high signal energy $\featEnergyCh{\idxCh}[\idxFrame]$, \ie, the observation noise of the \gls{kf} is inversely related to the signal energy $\featEnergyCh{\idxCh}[\idxFrame]$.

With all components of the \gls{kf} in place, the update equations for mean vector and state covariance are \cite{bishop_pattern_2006}
\begin{align}
\KFstateVec^{(\idxFeat)}[\idxFrame+1] &= \KFstateVec^{(\idxFeat)}[\idxFrame] + \KFgain^{(\idxFeat)}[\idxFrame] \left(\KFobservationVec^{(\idxFeat)}[\idxFrame] - \KFstateVec^{(\idxFeat)}[\idxFrame]\right),
\label{eq:KF-update-state} \\
\KFpriorStateCovarianceMat^{(\idxFeat)}[\idxFrame+1] &= %
\left( \KFpriorStateCovarianceMat^{(\idxFeat)}[\idxFrame] + \KFprocessNoiseCovarianceMat \right) %
\left(\mat{I}_{\numStates} -  \KFgain^{(\idxFeat)}[\idxFrame]\right), %
\label{eq:KF-update-state-covariance}
\end{align}
with the Kalman gain matrix
\begin{align}
\KFgain^{(\idxFeat)}[\idxFrame] &= \left(\KFpriorStateCovarianceMat^{(\idxFeat)}[\idxFrame]  + \KFprocessNoiseCovarianceMat \right) %
\left(\KFpriorStateCovarianceMat^{(\idxFeat)}[\idxFrame] + \KFprocessNoiseCovarianceMat + \KFobservationNoiseCovarianceMat[\idxFrame]\right)^{-1} %
\label{eq:KF-update-gain}
\end{align}
and initial values
\begin{align}
\KFstateVec^{(\idxFeat)}[0] &= \mathbf{0}_{\numStates},
&
\KFpriorStateCovarianceMat^{(\idxFeat)}[0] &= \mat{I}_{\numStates}.
\label{eq:KF-initial-values}
\end{align}
Note that the updates in \cref{eq:KF-update-state,eq:KF-update-state-covariance,eq:KF-update-gain} can be computed very efficiently since all involved matrices are diagonal.

For each time frame, after updating the \glspl{kf} for all features $\idxFeat \in \{1,\ldots,\numFeat\}$, the elements of the covariance matrix $\featCorrRawMat^{(\idxFeat)}[\idxFrame]$ are recovered from the \gls{kf} mean vector $\KFstateVec^{(\idxFeat)}[\idxFrame]$ by reversing the half-vectorization, \ie,
\begin{equation}
\hspace*{-2mm}
\featCorrRawCh{\idxCh,\idxChAlt}^{(\idxFeat)}[\idxFrame] %
= \left[\featCorrRawMat^{(\idxFeat)}[\idxFrame]\right]_{\idxCh\idxChAlt} %
= \left[\vech^{-1} \left( \KFstateVec^{(\idxFeat)}[\idxFrame] \right)\right]_{\idxCh\idxChAlt}.
\label{eq:KF-state-inversevech}
\end{equation}

Finally, the per-feature cross-channel \glspl{pcc} as elements of a per-feature \gls{pcc} matrix $\featCorrMat^{(\idxFeat)}[\idxFrame]$ are obtained from the estimated covariances by element-wise normalization
\begin{align}
\featCorrCh{\idxCh,\idxChAlt}^{(\idxFeat)}[\idxFrame] %
= \left[\featCorrMat^{(\idxFeat)}[\idxFrame]\right]_{\idxCh\idxChAlt} %
= \frac{\featCorrRawCh{\idxCh,\idxChAlt}^{(\idxFeat)}[\idxFrame]}%
{\sqrt{\featCorrRawCh{\idxCh,\idxCh}^{(\idxFeat)}[\idxFrame] \cdot \featCorrRawCh{\idxChAlt,\idxChAlt}^{(\idxFeat)}[\idxFrame]}}.
\label{eq:feature-correlation-coefficients}
\end{align}

\textit{Step 2) Feature Combination:}
As outlined earlier, the \gls{pcc} matrices of different features $\featCorrMat^{(\idxFeat)}[\idxFrame]$ capture different aspects of the underlying inter-channel coherence.
To recover an estimate of the inter-channel coherence from the multiple feature correlation coefficient matrices, we consider channel-wise matrices
\begin{align}
\featChanMat_{\idxCh}[\idxFrame] %
= \begin{bmatrix}
\featCorrCh{\idxCh,1}^{(1)}[\idxFrame] & \ldots &  \featCorrCh{\idxCh,1}^{(\numFeat)}[\idxFrame] \\
\vdots &  & \vdots \\
\featCorrCh{\idxCh,\numCh}^{(1)}[\idxFrame] & \ldots & \featCorrCh{\idxCh,\numCh}^{(\numFeat)}[\idxFrame]
\end{bmatrix} \in \realNumbers^{\numCh\times\numFeat},
\label{eq:similarity-perchannel}
\end{align}
where each $\featChanMat_{\idxCh}[\idxFrame]$ contains the inter-channel \glspl{pcc} of all $\numFeat$ feature sequences of all $\numCh$ microphone channels \wrt the corresponding feature sequence of a reference channel $\idxCh$.

Note that each column of $\featChanMat_{\idxCh}[\idxFrame]$, corresponding to one particular signal feature, models the same underlying inter-channel coherence.
The \glspl{pcc} of different features are then combined for each channel $\idxCh$ by extracting the dominant column structure of $\featChanMat_{\idxCh}[\idxFrame]$, \ie, finding its best rank-1 approximation in the \gls{ls} sense \cite{golub_matrix_2013}
\begin{align}
\min_{\featChanSingVal_{\idxCh}[\idxFrame],\featChanLeftSingVec_{\idxCh}[\idxFrame],\featChanRightSingVec_{\idxCh}[\idxFrame]} %
\left\lVert
\featChanMat_{\idxCh}[\idxFrame] - \featChanSingVal_{\idxCh}[\idxFrame] \featChanLeftSingVec_{\idxCh}[\idxFrame]\featChanRightSingVec_{\idxCh}^{\decoTran}[\idxFrame]
\right\rVert^{2}_{2}.
\label{eq:svd-rank-1}
\end{align}
Since $\featChanMat_{\idxCh}[\idxFrame]$ is generally non-square, the solution of \cref{eq:svd-rank-1} is obtained by the \gls{svd}, where $\featChanSingVal_{\idxCh}[\idxFrame] \in \realNumbers^{+}$ is the largest singular value of $\featChanMat_{\idxCh}[\idxFrame]$, and $\featChanLeftSingVec_{\idxCh}[\idxFrame] \in \realNumbers^{\numCh}$ and $\featChanRightSingVec_{\idxCh}[\idxFrame] \in \realNumbers^{\numFeat}$ are the principal left and right singular vectors, respectively.
The principal left singular vector $\featChanLeftSingVec_{\idxCh}[\idxFrame]$ captures the contribution of each channel to the dominant structure of $\featChanMat_{\idxCh}[\idxFrame]$, while the principle right singular vector $\featChanRightSingVec_{\idxCh}[\idxFrame]$ captures the contribution of each feature to the dominant structure.

To facilitate tracking of $\featChanLeftSingVec_{\idxCh}[\idxFrame]$ in time-variant scenarios and avoid recomputation of the full \gls{svd} in each time step, the principal left singular vector is instead iteratively refined over time.
To this end, recall that the left singular vectors of $\featChanMat_{\idxCh}[\idxFrame]$ are identical to the eigenvectors of the Gram matrix $\featChanMat_{\idxCh}[\idxFrame] \featChanMat_{\idxCh}^{\decoTran}[\idxFrame]$ \cite{golub_matrix_2013}.
Thus, given an estimate from the previous time step, the principal singular vector can be estimated using power methods \cite{golub_matrix_2013} as
\begin{align}
\Check{\featChanLeftSingVec}_{\idxCh}[\idxFrame+1] &= \left(\featChanMat_{\idxCh}[\idxFrame+1] \featChanMat_{\idxCh}^{\decoTran}[\idxFrame+1]\right) \featChanLeftSingVec_{\idxCh}[\idxFrame], %
    \label{eq:PI-matmult} \\
\featChanLeftSingVec_{\idxCh}[\idxFrame+1] &= \frac{\Check{\featChanLeftSingVec}_{\idxCh}[\idxFrame+1]} %
{\lVert \Check{\featChanLeftSingVec}_{\idxCh}[\idxFrame+1] \rVert_2}. %
\label{eq:PI-norm}
\end{align}
Note that the spectrum of $\featChanMat_{\idxCh}[\idxFrame]$ varies slowly over time such that a single iteration of \cref{eq:PI-matmult,eq:PI-norm} is sufficient.

In order to restore the intuitive notion of a similarity measure, the estimated principal singular vectors from \cref{eq:PI-norm} are re-normalized such that the similarity of each channel to itself is equal to one, and then concatenated to form the overall channel similarity matrix %
\begin{align}
\simMat[\idxFrame] = \begin{bmatrix} \frac{\featChanLeftSingVec_{1}[\idxFrame]}{\left[\featChanLeftSingVec_{1}[\idxFrame]\right]_1}, %
& \ldots, %
& \frac{\featChanLeftSingVec_{\numCh}[\idxFrame]}{\left[\featChanLeftSingVec_{\numCh}[\idxFrame]\right]_{\numCh}} \end{bmatrix} \in \realNumbers^{\numCh\times\numCh}.
\label{eq:similarity-matrix}
\end{align}

\textit{Step 3) Spectral Graph Partitioning:}
Microphone selection is equivalent to partitioning the set of available microphones $\setAllCh$ into two, potentially time-variant, disjoint subsets comprising the selected and discarded microphones, respectively.
Recalling the signal model \cref{eq:signal-model}, we use the convention that the former subset $\setSelCh[\idxFrame]$ contains the microphones capturing the \gls{soi} with high quality while the latter subset is its complement $\overline{\setSelCh}[\idxFrame]$ for those microphones dominated by the non-coherent noise field.
Relaxing the hard assignment of microphones to these subsets to a soft assignment leads to continuous real-valued utility estimates as shown in the following.
Spectral partitioning techniques \cite{chung_spectral_1997,fiedler_algebraic_1973,von_luxburg_tutorial_2007} operating on graph structures can determine such optimal partitionings very efficiently, especially when the number of microphones $\numCh$ is large.

Thus, we model the pairwise similarity of microphone channels using a time-variant graph structure $\graph(\graphVertices,\graphEdges[\idxFrame])$ \cite{chung_spectral_1997}, comprising a set of vertices $\graphVertices$ representing microphones and a set of weighted edges $\graphEdges[\idxFrame]$ representing the microphones' similarity at time frame $\idxFrame$.
For each edge $(\idxCh,\idxChAlt,\graphEdgeWeight_{\idxCh\idxChAlt}[\idxFrame]) \in \graphEdges[\idxFrame]$, the weight $\graphEdgeWeight_{\idxCh\idxChAlt}[\idxFrame] \in [0,1]$ captures the similarity of microphones $\idxCh$ and $\idxChAlt$.
The graph is equivalently specified by its weighted adjacency matrix $\matGraphConn[\idxFrame] \in \realNumbers^{\numCh \times \numCh}$, containing all weights $\graphEdgeWeight_{\idxCh\idxChAlt}[\idxFrame], \forall \idxCh, \idxChAlt \in \setAllCh$.
The pairwise microphone similarity should be a symmetric measure, \ie, channel $\idxCh$ should be as similar to $\idxChAlt$ as channel $\idxChAlt$ is to $\idxCh$, such that $\graphEdgeWeight_{\idxCh\idxChAlt}[\idxFrame] = \graphEdgeWeight_{\idxChAlt\idxCh}[\idxFrame]$.
To reflect this symmetry and the varying degrees of similarity, the graph should be undirected and weighted.
Since the matrix $\simMat[\idxFrame]$ in \cref{eq:similarity-matrix} does not necessarily exhibit these properties, only the symmetric part of its element-wise magnitude is used to construct the weighted adjacency matrix $\matGraphConn[\idxFrame]$, \ie,
\begin{align}
\graphEdgeWeight_{\idxCh\idxChAlt}[\idxFrame] = 
\left[\strut\matGraphConn[\idxFrame]\right]_{\idxCh\idxChAlt} %
= \frac{1}{2} \left( \left|\left[\strut\simMat[\idxFrame]\right]_{\idxCh\idxChAlt}\right| + \left|\left[\strut\simMat[\idxFrame]\right]_{\idxChAlt\idxCh}\right| \right).
\label{eq:graph-connectivity-matrix}
\end{align}

The \emph{degree} \cite{chung_spectral_1997} of the $\idxCh$-th vertex is defined as the sum of all outgoing edges' weights
\begin{align}
\graphDegree_{\idxCh}[\idxFrame] &= \sum_{\idxChAlt=1}^{\numCh} \graphEdgeWeight_{\idxCh\idxChAlt}[\idxFrame],
\label{eq:node-degree}
\end{align}
which are collected in the diagonal degree matrix
\begin{align}
\matGraphDegree[\idxFrame] &= \Diag \left\{\graphDegree_{1}[\idxFrame], \ldots, \graphDegree_{\numCh}[\idxFrame]\right\}.
\label{eq:mat-degree}
\end{align}
Note that $\graphDegree_{\idxCh}[\idxFrame] \geq 1, \forall\idxCh\in\setAllCh$ since the sum in \cref{eq:node-degree} includes $\graphEdgeWeight_{\idxCh\idxCh}[\idxFrame]=1$, which ensures invertibility of $\matGraphDegree[\idxFrame]$ even for degenerate graphs.

For an ideal partitioning, like for clustering, it is desirable that microphone signals belonging to the same group are similar while microphone signals belonging to different groups are dissimilar to allow for a clear distinction between the selected and the discarded microphones.

Using \cref{eq:signal-model} gives an interpretation in the context of microphone selection: \gls{soi}-dominated microphones exhibit strongly mutually correlated feature sequences and thus form one of the two partition subsets, while the feature sequences of noise-dominated microphones are only weakly correlated with the \gls{soi}-dominated microphones, and thus form the other subset.
In addition, even if the noise components $\sigNoise_{\idxCh}[\idxTime]$ are uncorrelated, their features likely are correlated, especially if they capture underlying statistics like variance.
These inter-group and intra-group similarities of a set $\setSelCh[\idxFrame] \subset \setAllCh$ and its complement $\overline{\setSelCh}[\idxFrame]$ are measured by \cite{von_luxburg_tutorial_2007}
\begin{align}
\cut(\setSelCh[\idxFrame], \overline{\setSelCh}[\idxFrame]) &= \sum_{\mathclap{\idxCh \in \setSelCh[\idxFrame], \idxChAlt \in \overline{\setSelCh}[\idxFrame]}} \graphEdgeWeight_{\idxCh \idxChAlt}[\idxFrame],
\label{eq:volume-cut} \\
\vol(\setSelCh[\idxFrame]) &= \sum_{\idxCh \in \setSelCh[\idxFrame]} \graphDegree_{\idxCh}[\idxFrame],
\label{eq:volume-set}
\end{align}
respectively.
Balancing the inter- and intra-group similarity to avoid degenerate solutions yields the \emph{normalized cut} objective function \cite{shi_normalized_2000}
\begin{align}
\ncut(\setSelCh[\idxFrame],\overline{\setSelCh}[\idxFrame]) &= %
\cut(\setSelCh[\idxFrame],\overline{\setSelCh}[\idxFrame]) \, \cdot \nonumber \\
&\hphantom{=}\; 
\left(
\frac{1}{\vol(\setSelCh[\idxFrame])} + %
\frac{1}{\vol(\overline{\setSelCh}[\idxFrame])}
\right).
\label{eq:objective-ncut}
\end{align}
As shown in \cite{shi_normalized_2000}, minimization of \cref{eq:objective-ncut} \wrt $\setSelCh$ can be reformulated as minimization of the generalized Rayleigh quotient
\begin{equation}
\hspace*{2pt}
\scalebox{0.96}{$\displaystyle
\min_{\mathclap{\hspace*{2mm}\setSelCh[\idxFrame],\overline{\setSelCh}[\idxFrame]}} \; \ncut(\setSelCh[\idxFrame],\overline{\setSelCh}[\idxFrame]) = %
\min_{\indicatorVec[\idxFrame]} \frac{\indicatorVec^{\decoTran}[\idxFrame] \left( \matGraphDegree[\idxFrame] - \matGraphConn[\idxFrame] \right) \indicatorVec[\idxFrame]}{\indicatorVec^{\decoTran}[\idxFrame] \matGraphDegree[\idxFrame] \indicatorVec[\idxFrame]},
$}
\label{eq:ncut-rayleigh-equivalence}
\end{equation}
where $\indicatorVec[\idxFrame]$ is a $\numCh$-dimensional discrete indicator vector satisfying
\begin{equation}
\indicatorVec^{\decoTran}[\idxFrame] \matGraphDegree[\idxFrame] \mathbf{1}_{\numCh} = 0.
\label{eq:ncut-orth-constraint}
\end{equation}
Additionally, the elements of $\indicatorVec[\idxFrame]$ may only take either of two values \cite{shi_normalized_2000}
\begin{equation}
\left[\vphantom{\big(}\indicatorVec[\idxFrame]\right]_{\idxCh} \in \left\{
    1,\; 
    \frac{\sum_{\idxChAlt \in \setSelCh[\idxFrame]} \graphDegree_{\idxChAlt}[\idxFrame]}{\sum_{\idxChAlt \in \overline{\setSelCh}[\idxFrame]} \graphDegree_{\idxChAlt}[\idxFrame]} \right\}.
\label{eq:ncut-discrete-constraint}
\end{equation}

When the discreteness constraint \cref{eq:ncut-discrete-constraint} on $\indicatorVec[\idxFrame]$ is relaxed to allow arbitrary real values, \ie, $\indicatorVec[\idxFrame] \in\realNumbers^{\numCh}$, the minimizer of the generalized Rayleigh quotient in \cref{eq:ncut-rayleigh-equivalence} is a solution to the generalized eigenvalue problem 
\begin{equation}
\left(\matGraphDegree[\idxFrame] - \matGraphConn[\idxFrame]\right) \indicatorVec[\idxFrame] = \lambda[\idxFrame] \matGraphDegree[\idxFrame] \indicatorVec[\idxFrame],
\label{eq:ncut-generalized-ev}
\end{equation}
where $\lambda[\idxFrame]$ is the generalized eigenvalue and $\indicatorVec[\idxFrame]$ is the generalized eigenvector.
The equivalent standard eigenvalue problem is obtained by left-multiplication of $\matGraphDegree^{-1}[\idxFrame]$
\begin{equation}
\matGraphLaplacian[\idxFrame] \indicatorVec[\idxFrame] = \lambda[\idxFrame] \indicatorVec[\idxFrame]
\label{eq:ncut-standard-ev}
\end{equation}
with the normalized random-walk Laplacian matrix \cite{chung_spectral_1997}
\begin{align}
\matGraphLaplacian[\idxFrame] &= \matGraphDegree^{-1}[\idxFrame] \left(\matGraphDegree[\idxFrame] - \matGraphConn[\idxFrame]\right) \nonumber \\
&=\mat{I}_{\numCh} - \matGraphDegree^{-1}[\idxFrame] \matGraphConn[\idxFrame].
\label{eq:mat-laplacian}
\end{align}
Thus, an approximate minimizer of \cref{eq:ncut-rayleigh-equivalence} is obtained by finding the smallest eigenvalue and its corresponding eigenvector of $\matGraphLaplacian[\idxFrame]$.
The trivial eigenvalue $0$ and its corresponding eigenvector $\vect{1}_{\numCh}$ \cite{von_luxburg_tutorial_2007} is excluded by the constraint \cref{eq:ncut-orth-constraint}.
Thus, the solution is the so-called \emph{Fiedler vector} $\vecFiedler[\idxFrame]$, \ie, the eigenvector corresponding to the smallest non-trivial eigenvalue of $\matGraphLaplacian[\idxFrame]$ \cite{von_luxburg_tutorial_2007}, which automatically satisfies \cref{eq:ncut-orth-constraint} as shown in \cite{shi_normalized_2000}.
While an approximate solution to the discrete problem can be obtained by discretizing $\vecFiedler[\idxFrame]$, \eg, based on the sign of each element, here we use the real-valued solution directly as an estimate of the microphones' utility.

As an eigenvector, the scale and in particular the sign of $\vecFiedler[\idxFrame]$ is ambiguous, \ie, both $-\vecFiedler[\idxFrame]$ and $\vecFiedler[\idxFrame]$ are valid solutions to the eigenvalue problem \cref{eq:ncut-standard-ev}.
The same holds for the objective function \cref{eq:objective-ncut}, which is invariant to exchanging $\setSelCh[\idxFrame]$ with $\overline{\setSelCh}[\idxFrame]$.
This ambiguity is usually not a problem for partitioning, since only the association of vertices to groups is desired, but not the identity of each group.
In other words, the partitioning given by $\vecFiedler[\idxFrame]$ distinguishes between the most and least useful microphones, but does not say which group is which.
Additionally, in low-\gls{snr} scenarios, noise-dominated microphone signals may exhibit large feature \gls{pcc} values due to similar noise signal statistics despite only weakly coherent noise signals.
To facilitate this distinction, supplemental side information, captured by the vector $\vecSupplemental[\idxFrame]$, is correlated with the preliminary utility estimates
\begin{align}
\pccUtilityEntropy[\idxFrame] = \pearsonCorr\left( \vphantom{\big(} \vecFiedler[\idxFrame], \vecSupplemental[\idxFrame] \right).
\end{align}
Depending on the sign of the Pearson Correlation Coefficient $\pccUtilityEntropy[\idxFrame]$, the sign of the estimated utility values is flipped to produce the final utility estimates
\begin{align}
\vecUtility[\idxFrame] &= \begin{cases}
\hphantom{-}\vecFiedler[\idxFrame] & \text{if } \pccUtilityEntropy[\idxFrame] \geq 0 \\
-\vecFiedler[\idxFrame] & \text{if } \pccUtilityEntropy[\idxFrame] < 0 \\
\end{cases}.
\label{eq:utility-resolution}
\end{align}

In \cite{gunther_network-aware_2021}, the supplemental information was chosen as the node degree, \ie, $\left[\vecSupplemental[\idxFrame]\right]_{\idxCh} = \graphDegree_{\idxCh}[\idxFrame]$.
While this choice allows detection of outliers if the volumes of the two subsets in the partition are very different, \ie, a large majority of microphones is either useful or not useful, it also requires further assumptions or knowledge about the identity of the majority group, \eg, that the majority of microphones observes the desired \gls{soi}.
To address these shortcomings, we consider typical \gls{soi} and interfering signals:
typical \gls{soi} signals, like speech, exhibit spectro-temporal structure while typical signal degradations, like reverberation or additive non-coherent noise, reduce said structure.
Thus, the differential signal entropy \cite{cover_elements_2006} is used to capture the structuredness of the observed signals
\begin{align}
\left[\vphantom{\big(}\vecSupplemental[\idxFrame]\right]_{\idxCh} %
&= - \entropy \left( \sigBlockCh{\idxCh}[\idxFrame]\right) \nonumber \\
&= \sum_{\mathclap{\idxHistBin = 0}}^{\numHistBin-1} %
    \estimate{\probabilityDensity}(\idxHistBin; \idxFrame) \; %
     \log_{2}(\estimate{\probabilityDensity}(\idxHistBin; \idxFrame))
\label{eq:differential-entropy}
\end{align}
as in \cite{gunther_online_2021}.
Therein, the \gls{pdf} is estimated by its $\numHistBin$-bin histogram
\begin{equation}
\estimate{\probabilityDensity}(\idxHistBin; \idxFrame) = \frac{1}{\blockLength} \left|\left\{ \idxTime,\; %
\histBinEdge_{\idxHistBin} \leq %
\left[\strut\sigBlockCh{\idxCh}[\idxFrame]\right]_{\idxTime} < %
\histBinEdge_{\idxHistBin+1} \right\}\right|
\label{eq:histogram-probability}
\end{equation}
with $\histBinEdge_{\idxHistBin}$ denoting the histogram bin edges.
Note that, for the experiments conducted in \cref{sec:experiments}, the signal blocks used to estimate entropy in \cref{eq:differential-entropy} are chosen longer than those for the feature extraction.
The entire microphone utility estimation procedure using Spectral Graph Partitioning is concisely summarized as pseudocode in \cref{alg:online-estimation}.

In the presence of point-like interferers, the signal model \cref{eq:signal-model} no longer strictly holds, such that it should be understood as a first step towards developing methods for more general acoustic scenarios,
Hence, somewhat degraded estimation performance must be expected, although the extent of degradation depends on the particular scenario.
For example, in an \gls{asn} spanning to rooms each with their own \gls{soi} with only low-level cross-talk between rooms and low-level additive noise, groups of useful microphones for either \gls{soi} can be identified, which still matches well with the desired outcome.
As a second example, consider an \gls{asn} in a single room, with two closely-spaced point sources.
For temporally overlapping source activity with both sources contributing similar signal power to each microphone, all microphones exhibit reduced utility \wrt either source as the other source is considered as noise, again matching qualitatively with reduced feature coavariance.
For source counting and associating the microphone subsets with the correct \gls{soi}, additional mechanisms need to be developed that are beyond the scope of this paper.

\begin{algorithm}[t]
\caption{Recursive microphone utility update using Spectral Graph Partitioning}
\label{alg:online-estimation}
\begin{algorithmic}
\REQUIRE $\featChanLeftSingVec_{\idxCh}[\idxFrame-1],\quad\forall\idxCh\in\setAllCh$ (previous similarity vectors)
\REQUIRE $\featCorrCh{\idxCh,\idxChAlt}[\idxFrame],\quad\forall\idxCh,\idxChAlt\in\setAllCh, \idxFeat\in\{1,\ldots,\numFeat\}$
\ENSURE $\vecUtility[\idxFrame]$ (estimated utility vector)

\FOR{microphones $\idxCh=1$ to $\numCh$}
\STATE $\featChanLeftSingVec_{\idxCh}[\idxFrame]$ $\leftarrow$ updated similarity vector, see \cref{eq:similarity-perchannel,eq:PI-matmult,eq:PI-norm}
\ENDFOR

\STATE $\simMat[\idxFrame]$ $\leftarrow$ concatenated normalized $\featChanLeftSingVec_{\idxCh}[\idxFrame]$, $\forall\idxCh\in\setAllCh$, see \cref{eq:similarity-matrix}
\STATE $\matGraphConn[\idxFrame]$ $\leftarrow$ symmetric adjacency matrix, see \cref{eq:graph-connectivity-matrix}
\STATE $\matGraphLaplacian[\idxFrame]$ $\leftarrow$ random-walk graph Laplacian, see \cref{eq:mat-laplacian}
\STATE $\vecFiedler[\idxFrame]$ $\leftarrow$ Fiedler vector of $\matGraphLaplacian[\idxFrame]$, see \cref{eq:mat-laplacian}
\STATE $\vecUtility[\idxFrame]$ $\leftarrow$ $\pm\vecFiedler[\idxFrame]$, after disambiguation \cref{eq:utility-resolution}

\end{algorithmic}
\end{algorithm}

\subsection{Importance of Specific Signal Features}
\label{sec:feature-importance}
Choosing an appropriate set of characteristic signal features for the microphone signals is vital:
too few features result in low estimation accuracy, while too many features unnecessarily strain the wireless network.
Even for an appropriate number of features, inappropriate features may even reduce overall estimation accuracy.
To explore the importance of individual signal features, we cast the feature selection as a \gls{ls} regression problem with a sparsity-promoting regularizer in order to obtain a low regression error while using as few features as possible.
Specifically, we interpret the matrix $\featChanMat_{\idxCh}[\idxFrame]$ as a dictionary matrix whose columns, or \emph{atoms}, contain the cross-channel correlation coefficients between the reference channel $\idxCh$ and all channels for one specific signal feature, and which are linearly combined to approximate the \gls{msc} of the observed microphone signals.
However, for the purpose of estimating microphone utility and microphone selection, the relative utility of microphone channels is more important than the absolute values, such that the zero-mean \gls{msc} vector, \ie,
\begin{align}
\vecMSCZeroMean[\idxFrame] &= \vecMSC[\idxFrame] - \left( \frac{1}{\numCh}\sum_{\idxCh=1}^{\numCh} \msc_{\idxCh}[\idxFrame] \right) \mathbf{1}_{\numCh},
\label{eq:msc-vec-zeromean)}
\end{align}
is used as the target quantity.
Thus, the $\ell_1$-regularized \gls{ls} cost function for a single acoustic scenario comprising $\numCh$ microphone signals with $\numFrame$ time frames is
\begin{equation}
\costWeightFeature(\vecWeightFeature) =
 \frac{1}{\numCh\numFrame} \sum_{\idxCh=1}^{\numCh} \sum_{\idxFrame=1}^{\numFrame} \lVert \vecMSCZeroMean[\idxFrame] - \featChanMat_{\idxCh}[\idxFrame] \vecWeightFeature \rVert_{2}^{2} 
+ \lassoWeight \lVert \vecWeightFeature \rVert_1,
\label{eq:cost-lasso}
\end{equation}
where $\vecWeightFeature = \begin{bmatrix} \weightFeatureBase_{1}, & \ldots, & \weightFeatureBase_{\numFeat} \end{bmatrix}^{\decoTran} \in \realNumbers^{\numFeat}$ captures the contribution of each feature and the parameter $\lassoWeight \in \realNumbers^{+}$ indirectly controls the sparsity of the vector, \ie, the number of used features.

Summing $\costWeightFeature(\vecWeightFeature)$ in \cref{eq:cost-lasso} over $\numTrialSim=120$ experiment trials (see \cref{sec:acoustic-setup-synth}) and then minimizing the sum yields the features weights $\vecWeightFeature$ depicted in \cref{fig:feature-weights}.
Naturally, higher values of $\lassoWeight$ result in sparser solutions, \ie, less selected features, ranging from 3 features to 12 features for the considered range of $\lassoWeight$.
The most important features appear to be lower-order statistical moments of the temporal waveform (\texttt{td\textunderscore{}centroid}, \texttt{td\textunderscore{}spread}, \texttt{td\textunderscore{}skewness}), higher-order statistical moments of the magnitude spectrum (\texttt{sd\textunderscore{}skewness}, \texttt{sd\textunderscore{}kurtosis}), and features capturing the temporal variation of the magnitude spectrum (\texttt{sd\textunderscore{}flux}, \texttt{sd\textunderscore{}variation}, \texttt{sd\textunderscore{}fluxnorm}).
For $\lassoWeight=0.001$, the selection comprises the four features \texttt{td\textunderscore{}skewness}, \texttt{sd\textunderscore{}slope}, \texttt{sd\textunderscore{}kurtosis}, and \texttt{sd\textunderscore{}fluxnorm}, two of which were part of the informal selection made in \cite{gunther_single-channel_2019}.
To keep the number of selected features similar to prior work \cite{gunther_network-aware_2021,gunther_online_2021}, we choose the aforementioned four features of $\lassoWeight=0.001$ for the experimental validation in \cref{sec:experimental-result}.

\section{Learning-based Utility Estimation}
\label{sec:learning-based}
\Glspl{ann} offer the ability to learn an optimum feature set (for given training data) to characterize the microphone signals, as well as optimally combining the features for estimating microphone utility.
Thus, we propose learning-based alternatives to both the model-based feature extraction (see \cref{sec:model-feature-extraction}) and the utility estimation (see \cref{sec:model-utility-estimation}) subsystems.
The extractor module in \cref{fig:architecture-extractor} realizes the feature extractor on the left-hand side of \cref{fig:system-overview} (both in red), while the estimator module in \cref{fig:architecture-estimator} realizes the utility estimator on the right-hand side of \cref{fig:system-overview} (both in blue).
For both subsystems in \cref{fig:system-overview}, the \gls{ann} architectures are chosen to reflect the modeling capabilities of their model-based counterparts.
Both modules are trained together in an end-to-end fashion.
During inference, the extractor and utility estimator modules run on the network nodes and the \gls{ap}, respectively, such that only the compressed feature representation need to be transmitted to the \gls{ap}.

\subsection{Node-wise Feature Extraction}
\label{sec:learning-feature-extraction}
The signal features discussed in \cref{sec:feature-importance}, although effective for utility estimation, are not necessarily optimally suited for utility estimation.
Learning a set of features specifically tailored to characterize microphone signals for the purpose of estimating utility promises improved accuracy and a more compact representation.
The structure of the feature extractor module is depicted schematically in \cref{fig:architecture-extractor}.
First, the magnitude spectrum of a single microphone signal block $\sigBlockCh{\idxCh}[\idxFrame]$ in \cref{eq:signal-block} is computed.
Due to the loss of phase information, this transform is not invertible and thus prevents the model from learning exact equivalents of the time-domain features in \cref{sec:model-feature-extraction}.
However, recalling that the ground-truth utility is given by the \gls{msc}, the magnitude spectrum appears to be an obvious choice for the representation of the input data.
In addition, models that use magnitude spectrum have outperformed their counterparts that use the time-domain waveform in our experiments.
The magnitude spectrum then passes through a series of fully connected feed-forward layers that get progressively narrower to condense information until a desired number of signal features is reached.

The final batch normalization and \gls{gru} layer allows the extractor module to learn features that describe the evolution of some quantity over time, \eg, spectral flux.
Trained weights are shared between the instances of the module at different microphones, \ie, no sensor-specific features are extracted.

\subsection{Utility Estimation}
\label{sec:learning-utility-estimation}
The architecture of the utility estimator is shown in \cref{fig:architecture-estimator} using the concatenated feature vectors $\featVecInstant_{\idxCh}[\idxFrame]$ from the individual microphones as an input.
The memory of the first \gls{gru} layer allows capturing the temporal evolution of the feature sequences.
The following fully connected layers all contain the same number of neurons and are responsible for regression of the \gls{gru} outputs onto the target \gls{msc} values.
Passing in feature sequences directly, as opposed to the \glspl{pcc} in the model-based method described in \cref{sec:model-based}, allows the network to differentiate between useful and non-useful microphones, such that no separate disambiguation step or supplemental information is needed.
Unlike the model-based estimation in \cref{sec:model-utility-estimation}, the number of microphones $\numCh$ directly determines the structure of the model which must be retrained when the number of microphones changes.

\section{Experimental Validation}
\label{sec:experiments}
The algorithms from \cref{sec:model-based,sec:learning-based} are evaluated on simulated and recorded acoustic data.
The considered scenarios feature both static and moving \glspl{soi}, different room dimensions and reverberation times, and different arrangements of $\numCh=10$ microphones, some of which may be physically obstructed by objects.
Estimation accuracy is quantified by computing the time-variant \gls{pcc} between the estimated utility vector $\vecUtility[\idxFrame]$ and the  \gls{msc} vector $\vecMSC[\idxFrame]$
\begin{equation}
\corrUtilityCoherence[\idxFrame] = \pearsonCorr(\vecUtility[\idxFrame], \vecMSC[\idxFrame]).
\label{eq:corr-utility-coherence}
\end{equation}

For the following experiments, signals are sampled at $\sampleFreq = \SI{16}{kHz}$.
For block processing, signals are partitioned into blocks of $\blockLength=1024$ samples with a block shift of $\blockShift=512$ samples.
As the only exception, differential entropy in \cref{eq:differential-entropy}, since it is estimated by a histogram approach, uses longer blocks of $32\,000$ samples for more robust estimates.
Due to the larger block size, the estimated differential entropy also changes more slowly over time, thus promoting temporal continuity of the estimated utility via \cref{eq:utility-resolution}.

For the proposed model-based approach from \cref{sec:model-based}, termed \methodModelProp in the following, the microphone signals are characterized using the four features identified in \cref{sec:feature-importance}, \ie, \texttt{td\textunderscore{}skewness}, \texttt{sd\textunderscore{}slope}, \texttt{sd\textunderscore{}kurtosis}, and \texttt{sd\textunderscore{}fluxnorm}.
The temporal recursive smoothing factor in \cref{eq:feature-vec-recursive} is chosen as $\featVecRecursionFactor = 0.99$, the scaling factors for the \gls{kf} process and observation noise are $\KFprocessNoiseCovarianceMatScale = 1$ and $\KFobservationNoiseCovarianceMatScale = 50$, respectively.

For the learning-based approach, the extractor contains 7 fully connected layers with $513, 256, 128, 64, 32, 16$ neurons, respectively, followed by a single \gls{gru} layer with 16 inputs and 16 hidden states.
Recall that $\blockLength=1024$ such that the $513$ inputs to the first layer correspond to the non-redundant part of the signal's magnitude spectrum.
The utility estimator contains a single \gls{gru} layer with $16\numCh = 160$ inputs and $10$ hidden states, followed by $3$ fully connected layers with $10$ neurons each.
Since identical copies of the extractor module are run for each microphone channel $\idxCh\in\setAllCh$, the total number of parameters is $175\,000$ for the extractor regardless of the number of microphones $\numCh$, and $5\,500$ for the utility estimator with the above configuration which scales asymptotically quadratically with the number of channels $\numCh$.
The objective function to be minimized in training is the \gls{mse} between the estimated utility $\vecUtility[\idxFrame; \setTrainableParameters]$ and the ground-truth \gls{msc} $\vecMSC[\idxFrame]$
\begin{equation}
\min_{\setTrainableParameters}  \frac{1}{\numFrame} \sum_{\idxFrame=1}^{\numFrame} \lVert \vecUtility[\idxFrame; \setTrainableParameters] - \vecMSC[\idxFrame] \rVert^{2}_{2},
\label{eq:learning-based-costfunction}
\end{equation}
where $\setTrainableParameters$ denotes the set of all trainable parameters.
The model is trained by the Adam optimizer \cite{kingma_adam_2017} with a learning rate of $10^{-3}$ on acoustic data split into 70\% training and 30\% validation data.
Due to the combinatorial construction of the acoustic data (see \cref{sec:acoustic-setup-synth,sec:acoustic-setup-real}), it is likely that the same speech signal occurs both in the training and the testing data.
However, they never occur in the same combination of source trajectory and simulated room which are the predominant influencing factors of microphone utility.

A total of 8 algorithmic variants and baselines are evaluated.
First, two baseline variants, termed \methodModelMSC and \methodModelCDR, use the cross-microphone \gls{msc} and the \gls{cdr} as their only respective feature.
The latter is computed by a \gls{doa}-independent estimator \cite{schwarz_coherent--diffuse_2015} assuming a diffuse noise coherence and has been successfully used in weighting and selecting observations made by different microphones \cite{brendel_stft_2018,gunther_microphone_2021}.
Note that although this implies oracle knowledge in the sense of signal availability at the \gls{ap}, the \gls{msc} is still computed from time-limited observation windows and thus entails all of the associated estimation challenges, \eg, \cite{carter_time_1981, carter_coherence_1987}.
The same holds for the \gls{cdr}, as it is based directly on the estimated \gls{msc}.
The model-based system described in \cref{sec:model-based} including the \glspl{kf} for covariance estimation is termed \methodModelProp.
To evaluate the effectiveness of the \gls{kf}, a variant of the proposed system is evaluated  that uses a simple recursive temporal smoothing like \cref{eq:feature-vec-recursive} for feature covariance estimation, termed \methodModelSmooth.
Furthermore, to judge the modeling capabilities of the \gls{ml}-estimator, \methodModelEstimatornet combines all 18 traditional features from \cref{sec:model-feature-extraction} and the \gls{ml}-based utility estimator module from \cref{sec:learning-utility-estimation}.
In addition, three different training variants are investigated:
The first variant \methodMLSynth uses exclusively simulated data.
For practical application, it is highly desirable to deploy a pre-trained model and fine-tune its parameters specifically toward a new unseen scenario.
To this end, a copy of the pre-trained \methodMLSynth is fine-tuned on recorded data, termed \methodMLTuned.
Finally, a third version trained on both simulated and recorded data simultaneously, termed \methodMLJoint, is also evaluated.

\subsection{Acoustic Data}
\label{sec:acoustic-setup}

\subsubsection{Simulated Data}
\label{sec:acoustic-setup-synth}
\begin{table}[h]
\centering
\begin{tabular}{ccccc}
\toprule
\multirow{2}{*}{Room} & \multicolumn{3}{c}{Dimensions} & \multirow{2}{*}{Reverberation Time $\reverbTime$} \\
& $x [\si{m}]$ & $y [\si{m}]$ & $z [\si{m}]$ & \\
\midrule
A & $5.0$ & $5.2$ & $3.0$ & \SI{500}{ms}  \\
B & $6.2$ & $5.0$ & $2.5$ & \SI{700}{ms}  \\
C & $4.8$ & $4.2$ & $2.3$ & \SI{350}{ms}  \\
\bottomrule
\end{tabular}
\caption{Dimensions and reverberation times of simulated rooms.}
\label{tab:setup-rooms}
\end{table}
\noindent
Microphone signals for a single \gls{soi} moving in a shoe box room are simulated using the image-source method \cite{allen_image_1979,habets_signal_2011}.
The \gls{soi} trajectory is spatially discretized to obtain a set of time-variant \glspl{rir} to be convolved with the \gls{soi} signal excerpts.
Male and female speech segments of \SI{28}{s} duration are used as \gls{soi} signals.
The source moves rapidly during the time intervals \SI{8}{s} -- \SI{10}{s} and \SI{18}{s} -- \SI{20}{s}, and otherwise moves slightly around a resting position to simulate the behaviour of human speakers.
Under these constraints, 20 different, random source trajectories are generated.
Three different rooms with typical living room-like acoustic properties (see \cref{tab:setup-rooms}) are considered.
In each room, $\numCh=10$ cardioid microphones are placed at random positions and with random azimuthal rotation.
In total, $\numTrialSim = 120$ distinct acoustic setups (20 trajectories $\times$ 2 signals $\times$ 3 rooms) are simulated, resulting in \SI{56}{min} of speech data.
The generated source images are superimposed with spatially uncorrelated white noise of an equal, fixed level to attain an \gls{snr} of \SI{10}{dB} at the microphone with the strongest source image on average.
Due to the lower \gls{soi} contribution, other microphones have a lower average \gls{snr}.
\Cref{fig:setup-synthetic} illustrates Room A along with an exemplary source trajectory.

\subsubsection{Recorded Data}
\label{sec:acoustic-setup-real}
The recorded acoustic data is obtained from $\numCh=10$ microphones arranged pair-wise in a quarter circle around a static loudspeaker representing the \gls{soi} as shown in \cref{fig:setup-audiolab}.
Although the microphones capsules are omnidirectional, they exhibhit nonuniform directivity due to being mounted in metal enclosures facing the \gls{soi} which causes diffraction.
\Gls{soi} signals comprise male, female and children's speech.
Instead of a moving source, different usefulness of the microphones is induced by occluding some of the sensors.
Obstacles may cover two microphone pairs as indicated in \cref{fig:setup-audiolab}, or a single microphone pair.
Additionally, obstacles consist of different materials, \ie, solid wood, foam and cloth, such that sound can permeate through some of them.
In total, $\numTrialRec = 36$ distinct acoustic setups (12 obstructions $\times $ 3 signals) are recorded, resulting in \SI{36}{min} of speech data.
Like for simulated data, spatially uncorrelated white noise is added to the recorded microphone signals to achieve an \gls{snr} of \SI{10}{dB}.

\subsection{Experimental Results}
\label{sec:experimental-result}

\subsubsection{Utility Estimation for Simulated Data}
\label{sec:experimental-result-synth}
\Cref{fig:results-simulated} shows the median, as well as lower and upper quartile, of the \gls{pcc} $\corrUtilityCoherence[\idxFrame]$ across trials as a function of the time frame $\idxFrame$ for simulated data, which should ideally produce values close to 1.
Whenever the source moves (indicated by the grey-shaded vertical areas in \cref{fig:results-simulated}), the source-microphone distances suddenly change causing the observed rapid decrease of $\corrUtilityCoherence[\idxFrame]$.
Both baseline methods \methodModelMSC and \methodModelCDR achieve limited performance due to the high relative noise levels in the microphone signals and the time-limited observation windows impeding accurate estimation.
Both purely model-based variants \methodModelProp and \methodModelSmooth exhibit good steady-state performance with \glspl{pcc} around $0.9$ and quick initial convergence and reconvergence after source motion.
The \gls{kf} variant \methodModelProp achieves slightly better accuracy on average than \methodModelSmooth, and is more robust, e.g., visible at around \SI{4}{s} and \SI{13}{s}.
The trained \methodModelEstimatornet offers only very small improvements over \methodModelProp and \methodModelSmooth despite using all of the features, indicating that the four selected features for \methodModelProp and \methodModelSmooth are close to optimal for these scenarios.
The learning-based \methodMLSynth trained on matching data achieves very similar performance, trading a more consistent performance when the \gls{soi} does not move for a slightly slower reconvergence behaviour.
As expected, fine-tuning the \gls{ml} model using recorded data significantly degrades performance for simulated data, as shown by \methodMLTuned.
Finally, the \gls{ml} model with both simulated and recorded data from the beginning, \ie, \methodMLJoint, clearly outperforms all other considered methods, with only minor breakdowns and very fast recovery.
Interestingly, incorporating recorded data besides the simulated into the training procedure also improves performance on simulated data.
Convergence of all methods is very fast, reaching peak accuracy almost instantaneously after the \gls{soi} becomes active after an initial silence period of about \SI{1}{s}.

While the \gls{ml} models implicitly learning the temporal structure of the source movement might be a concern here, our experiments with random time intervals of \SI{4}{s} to \SI{12}{s} between two successive source movements have shown no noticeable degradation compared to fixed time intervals.

\subsubsection{Utility Estimation for Recorded Data}
\label{sec:experimental-result-real}
As for the simulated data, \cref{fig:results-recorded} shows the median and quartiles of $\corrUtilityCoherence[\idxFrame]$ across trials as a function of the time frame $\idxFrame$ for recorded data.
Since the \gls{soi} is static, the usefulness of microphones is predominantly influenced by their occlusion and no clear temporal structure can be discerned.
Both of the oracle baselines achieve consistent but limited performance with \glspl{pcc} between $0.6$ and $0.8$.
The advantage of \methodModelCDR may be attributed to the diffuse noise coherence model, which enhances the contrast between microphones since residual coherence is considered as noise, particularly in low-frequency regions.
While the median of both \methodModelProp and \methodModelSmooth reaches $0.9$ after about \SI{2}{s}, their performance degrades over the experiment duration. %
Note that convergence of these two model-based variants and the baselines is initially delayed by about \SI{1.5}{s} due to incorrect disambiguation of the utility estimates in \cref{eq:utility-resolution}, indicating opportunity for future improvements.
This is reinforced by \methodModelEstimatornet, which simultaneously avoids this initial delay and achieves significantly better performance.
Thus, \methodModelEstimatornet, which has access to all features, outperforming \methodModelProp suggests that the four selected features are suboptimal for the type of degradation encountered in the recorded data.
The relatively weak performance of \methodMLSynth with median values of around $0.5$ is unsurprising since the model was not trained using recorded data.
The performance of \methodMLTuned is only on par with \methodModelProp, indicating that adjustment of a pre-trained model is not as straightforward as anticipated, likely due to pre-training driving the \gls{ann} parameters to a local minimum that cannot be escaped easily by subsequent tuning.
Like for simulated data, \methodMLJoint outperforms other methods on recorded data, achieving almost ideal values extremely fast and consistently, \ie, at almost no spread.
Because they share their architecture and thus modeling capability, the advantage of \methodMLJoint over \methodMLTuned is due to the different training data, which matches the phenomenon that unrelated data improves performance, as it is also observed for simulated data (see \cref{fig:results-simulated}).

\subsubsection{Identification of a Single Most Useful Microphone}
\label{sec:experimental-result-selection}
\begin{table}[h]
\centering
\begin{tabular}{lcc}
\toprule
Algorithm variant & Synthetic data & Recorded data \\
\midrule
average pair-wise CDR \cite{schwarz_coherent--diffuse_2015} %
                            & 0.4269 & 0.1717  \\
\midrule
\methodModelProp            & 0.5465 & 0.1551  \\
\methodModelEstimatornet    & 0.5698 & 0.1277  \\
\methodMLJoint              & \textbf{0.5793} & \textbf{0.1940}  \\
\bottomrule
\end{tabular}
\caption{Fraction of time frames where the single most useful microphone is identified correctly.
The ground-truth selection is given by the microphone with the maximum oracle SNR \cref{eq:snr-channel}.}
\label{tab:result-selection-hitrate}
\end{table}
In addition to the accuracy of the estimated continuous-valued utilities, the capability of different algorithmic variants to correctly identify the microphone with the highest utility is investigated.
To this end, the channel-wise \gls{snr} in \cref{eq:snr-channel} is computed using oracle knowledge of the individual signal components, \ie, the \gls{soi} source image and the additive noise.
The microphone that maximizes \cref{eq:snr-channel} is considered the most useful, representing the ground truth in this experiment.
Note that, as the \gls{snr} changes over time, so does the identity of the best microphone.
For brevity, we restrict the investigation to the best performing model-based and \gls{ml}-based variants, \ie, \methodModelProp, \methodModelEstimatornet and \methodMLJoint.
For each variant, the microphone with the highest estimated utility is selected.
In the absence of a more directly comparable approach, the microphone with the highest average pair-wise \gls{cdr} is selected as a baseline.
Therein, the \gls{cdr} is computed using the \gls{doa}-independent estimator \cite{schwarz_coherent--diffuse_2015} and a diffuse noise coherence model as described in \cref{sec:experimental-result}.
Because the microphones are connected to separate network nodes, this \gls{cdr} baseline requires transmission of all microphone signals to one of the network nodes, which limits its practical applicability in \glspl{asn}.
As performance measure, we use the fraction of time frames in which the estimated identity of the most useful microphone coincides with the \gls{snr}-based ground truth.

The obtained results are shown in \cref{tab:result-selection-hitrate} separately for simulated and recorded data.
For simulated data, all proposed variants clearly outperform the \gls{cdr} baseline, with \methodMLJoint achieving the highest accuracy as expected from the previous results.
In the more challenging scenarios with recorded data, the overall accuracy decreases for all methods.
Although the \gls{cdr} baseline outperforms both \methodModelProp and \methodModelEstimatornet which use hand-crafted signal features, the \gls{ml}-based \methodMLJoint outperforms all off the remaining considered methods.
It must be reiterated that the \gls{cdr} baseline in \cref{tab:result-selection-hitrate} uses the microphone signal \gls{msc} as oracle knowledge, requiring transmission of all microphone signals.
In contrast, the proposed \methodModelProp, \methodModelEstimatornet and \methodMLJoint have no such limitations.

\subsubsection{Discussion}
\label{sec:experimental-result-discussion}
Let us summarize the previous \cref{sec:experimental-result-synth,sec:experimental-result-real,sec:experimental-result-selection} and point out implications for practical application.
Out of the \gls{ml}-based approaches, \methodMLJoint provides very good utility estimates but require recorded acoustic data for training.
Comparing the results of \methodMLTuned and \methodMLJoint, adaptation of a pre-trained network to new scenarios is not straightforward, likely requiring more sophisticated transfer learning techniques.
As major drawback for \glspl{asn} in realistic conditions, obtaining a sufficient amount of labeled training data for a variety of acoustic scenarios is difficult since the estimation of the \gls{msc} values necessary for training require prior transmission and potentially synchronization of all observed signals.
While this could be remedied, \eg, by network nodes with enough memory to buffer the signals before transmission, this problem is beyond the scope of this contribution.
Furthermore, the architecture of the utility estimator module explicitly depends on the number of microphones $\numCh$ and thus requires retraining whenever $\numCh$ changes, \eg, new \gls{asn} nodes are added.

In contrast, the model-based approach \methodModelProp has shown a more modest, yet robust, performance for both simulated and recorded data.
It is also blind, \ie, does not require knowledge of array geometries, acoustic meta parameters like reverberation time, and especially the number of microphones $\numCh$.
Thus, it can be straightforwardly deployed in different acoustic environments with the need to collect acoustic data to train or fine-tune the model.
For a real system, a model-based scheme can be used as initial solution to collect labeled training data, which can then be used to tailor an \gls{ml}-based model to the specific acoustic scenario of the training data.

\section{Conclusion}
In this contribution, we tackled microphone utility estimation for \glspl{asn}.
Specifically, we revisited model-based approaches and discussed the usefulness of specific features, with features describing temporal variations and higher-order statistical moments of the signals' magnitude spectra being the most useful overall.
Furthermore, we proposed alternative, machine learning-based realizations to learn an optimal feature set and utility estimator.
Experimental validation showed that both model- and \gls{ml}-based approaches are viable in principle with their own strengths and drawbacks.
The model-based approach is straightforwardly applied to \glspl{asn} with an arbitrary number of microphones $\numCh$, but is clearly outperformed by suitably trained \gls{ml} models.
In contrast, the \gls{ml}-based approaches, particularly \methodMLJoint, achieve superior performance if matching training data are available.

\begin{backmatter}

\section*{Acknowledgements}%
The authors thank Adhithyan Ramadoss for his help acquiring the recorded acoustic data used in the experimental study.

\section*{Funding}%
This work was funded by the Deutsche Forschungsgemeinschaft (DFG, German Research Foundation) -- 282835863 -- within the Research Unit FOR2457 ``Acoustic Sensor Networks''.

\section*{Abbreviations}%
\printglossary[type=\acronymtype,title={}]

\section*{Availability of data and materials}%
The datasets used and/or analysed during the current study are available from the corresponding author on reasonable request.

\section*{Ethics approval and consent to participate}%
Not applicable.

\section*{Competing interests}
The authors declare that they have no competing interests.

\section*{Consent for publication}%
Not applicable.

\section*{Authors' contributions}
MG designed the proposed systems, designed and conducted the experimental studies, analyzed their results, and drafted the manuscript.
AB co-designed the proposed systems and the experiments, and provided invaluable technical feedback on the manuscript draft.
WK helped with interpreting the results, and provided extensive feedback on the manuscript draft, especially regarding its clarity.
All authors read and approved the final manuscript.

\bibliographystyle{bmc-mathphys} %
\bibliography{references}      %

\section*{Figures}
\begin{figure}[h!]
\centering
\includegraphics{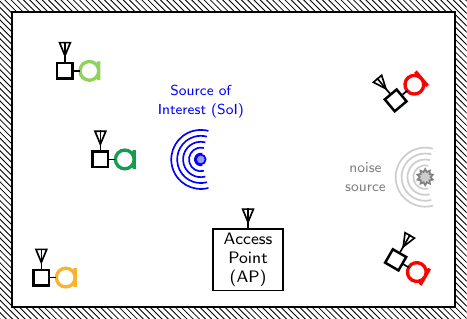}
\caption{Scenario for an ASN with a single SOI captured by spatially distributed microphones.}
\label{fig:introduction-scenario}
\end{figure}

\begin{figure*}[h!]
\centering
\includegraphics{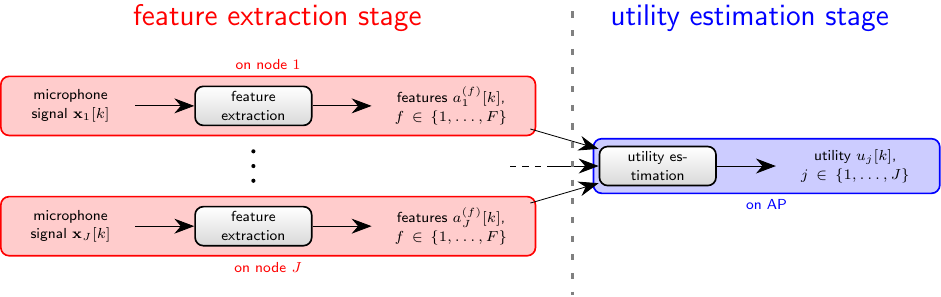}
\caption{System overview:
In the feature extraction stage, characteristic signal feature sequences are computed for each microphone signal independently.
Afterwards, the features sequences from all microphones are collected at the AP and used to estimate each microphone's utility.}
\label{fig:system-overview}
\end{figure*}

\begin{figure*}[h!]
\centering
\includegraphics{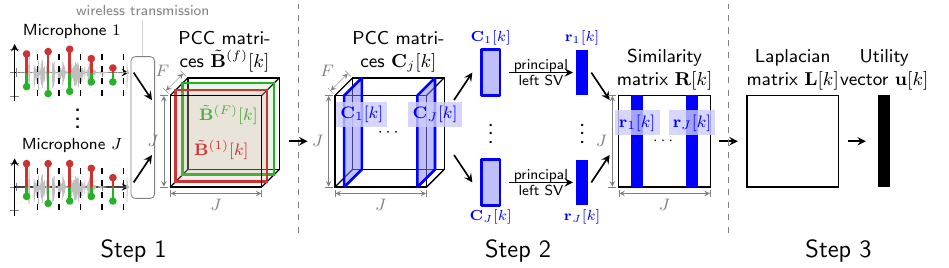}
\caption{Overview of model-based utility estimation.
For clarity, only two features are illustrated in Step 1.}
\label{fig:overview-model-utility-estimation}
\end{figure*}

\begin{figure}[h!]
\centering
\includegraphics{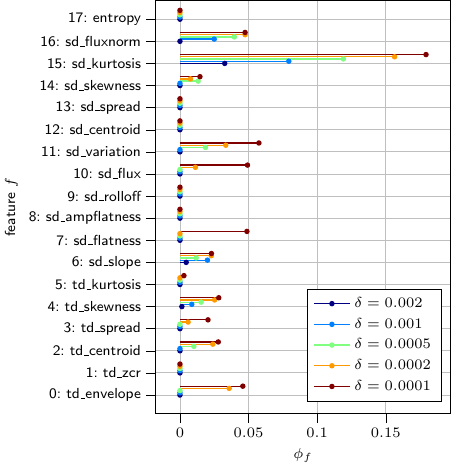}
\caption{Feature weights $\weightFeatureBase_{\idxFeat}$ for different values of $\lassoWeight$.}
\label{fig:feature-weights}
\end{figure}

\begin{figure}
\centering
\includegraphics{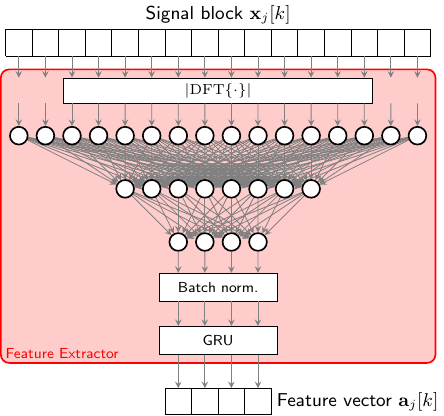}
\caption{Architecture of the Feature Extractor module.}
\label{fig:architecture-extractor}
\end{figure}

\begin{figure}[h!]
\centering
\includegraphics{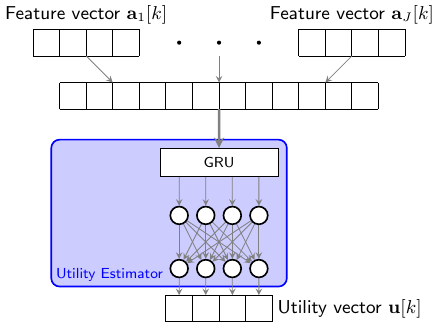}
\caption{Architecture of the Utility Estimator module.
Feature vectors from different microphones are concatenated to form a single, longer feature vector.
The GRUs exploit the temporal information contained in the feature sequences.
The FC layers estimate the microphone utility from the GRU outputs.}
\label{fig:architecture-estimator}
\end{figure}

\begin{figure}[h!]
\centering
\includegraphics{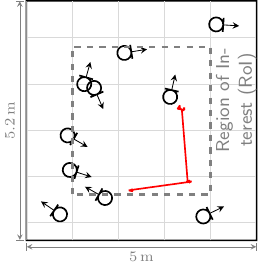}
\caption{Simulated room A and exemplary source trajectory (red) for synthesized data.}
\label{fig:setup-synthetic}
\end{figure}

\begin{figure}[h!]
\centering
\includegraphics{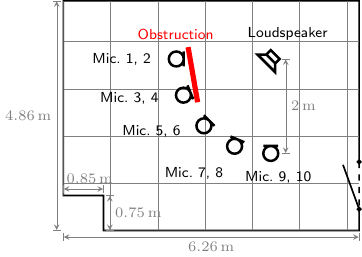}
\caption{Experiment setup and exemplary obstruction for recorded data.}
\label{fig:setup-audiolab}
\end{figure}

\begin{figure*}
\centering
\includegraphics{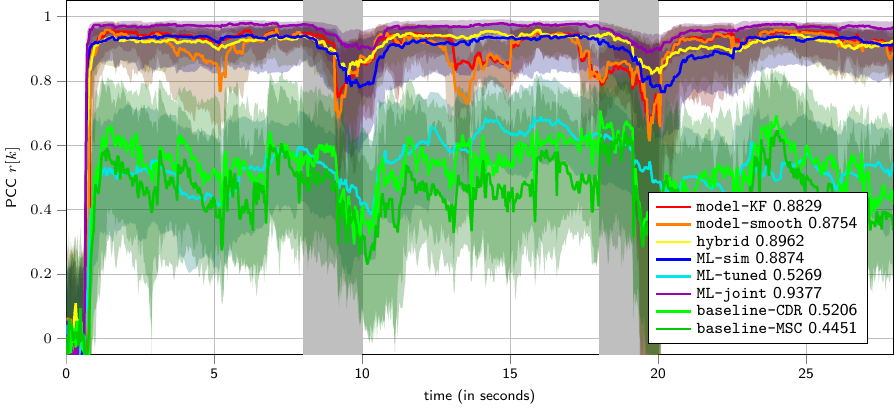}
\caption{Median (solid) and lower/upper quartile (shared areas) of PCC $\corrUtilityCoherence[\idxFrame]$ over all $\numTrialSim=120$ synthetic experiment trials.
Grey-shaded areas indicate time intervals of source movement.}
\label{fig:results-simulated}
\end{figure*}

\begin{figure*}
\centering
\includegraphics{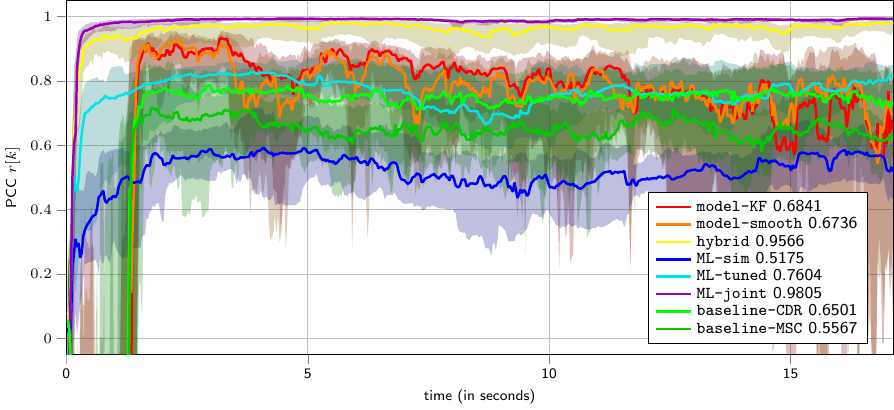}
\caption{Median (solid) and lower/upper quartile (shared areas) of PCC $\corrUtilityCoherence[\idxFrame]$ over all $\numTrialRec=36$ real-data experiment trials.}
\label{fig:results-recorded}
\end{figure*}

\section*{Tables}
All tables are placed within the manuscript as per submission guidelines due to their length of less than one A4 page.

\section*{Additional Files}
Not applicable.

\end{backmatter}
\end{document}